\documentclass[showpacs,
preprint,preprintnumbers,nofootinbib,
amsmath,amssymb]{revtex4}
\usepackage{color}
\usepackage{graphicx}
\usepackage{epsfig}

\allowdisplaybreaks

\let\jnfont=\rm
\def\NPB#1,{{\jnfont  Nucl.\ Phys.\ B }{\bf #1},}
\def\PLB#1,{{\jnfont Phys.\ Lett.\ B }{\bf #1},}
\def\EPJC#1,{{\jnfont Euro.\ Phys.\ J.\ C }{\bf #1},}
\def\PRD#1,{{\jnfont \em Phys.\ Rev.\ D }{\bf #1},}
\def\PRL#1,{{\jnfont Phys.\ Rev.\ Lett.\ }{\bf #1},}
\def\MPLA#1,{{\jnfont Mod.\ Phys.\ Lett.\ A }{\bf #1},}
\def\JPG#1,{{\jnfont J.\ Phys.\ G}{\bf #1},}
\def\CTP#1,{{\jnfont Commun.\ Theor.\ Phys.\ }{\bf #1},}
\def\CPC#1,{{\jnfont Chin. \ Phys. \ C}{\bf #1},}
\def\CPL#1,{{\jnfont Chin. \ Phys. \ Lett}{\bf #1},}
\def\RMP{\jnfont  Rev. Mod. Phys.}

\def\p_slash{\not{\hbox{\kern-2.1pt $p$}}}
\def\k_slash{\not{\hbox{\kern-2.1pt $k$}}}
\def\E_slash{\not{\hbox{\kern-5.1pt $E$}}}

\begin{document}
\title{PGB pair production at LHC and ILC as a probe of the
topcolor-assisted technicolor models}

\preprint{\parbox{1.2in}{\noindent arXiv:1105.2607 }}


\author{Guo-Li Liu$^1$\email{guoliliu@zzu.edu.cn}, Huan-Jun Zhang$^2$, Ping
Zhou$^{1,3}$  ~~ }

\affiliation{ $^1$Department of Physics, Zhengzhou University, Zhengzhou,
 China \\
  $^2$Department of Physics, Zhengzhou University of Light Industry, Zhengzhou,
 China \\
  $^3$Institut f\"{u}r Strahlenphysik, Forschungszentrum Dresden-Rossendorf, 01314~Dresden,
 Germany}

\begin{abstract}
The topcolor-assisted technicolor (TC2) model predicts some light
pseudo goldstone bosons (PGBs), which may be accessible at the LHC
or ILC. In this work we study the pair productions of the charged or
neutral PGBs at the LHC and ILC. For the productions at the LHC we
consider the processes proceeding through gluon-gluon fusion and
quark-antiquark annihilation, while for the productions at the ILC
we consider both the electron-positron collision and the
photon-photon collision. We find that in a large part of parameter
space the production cross sections at both colliders can be quite
large compared with the low standard model backgrounds. Therefore,
in future experiments these productions may be detectable and allow
for probing TC2 model.
\end{abstract}
\maketitle

\section{Introduction}
It is widely believed that the mechanism of electroweak symmetry
breaking(EWSB) and the origin of the particle mass remain prominent
mystery in current particle physics in spite of the success of the
standard model(SM) tested by high energy experimental data. There
has been no experimental evidence of the SM Higgs boson existing.
Furthermore, the neutrino oscillation experiments have made one
believe that neutrinos are massive, oscillate in flavor, which
presently provides the only experimental hints of new physics
\cite{neutrino}. Thus, the SM can only be an effective theory below
some high energy scales. Other EWSB mechanisms and extended Higgs
sectors have not been excluded in the theoretical point of view.

To completely avoid the problems arising from the elementary Higgs
scalar field in the standard model (SM), various kinds of models for
electroweak symmetry breaking (EWSB) have been proposed, among which
the technicolor models\cite{technicolor,tc2-rev} are attractive
because they provide a possible EWSB mechanism without introducing
an elementary scalar Higgs boson. In this kind of models, EWSB can
be achieved via introducing new strong interaction. Technicolor
models open up new possibilities for new physics beyond the SM,
which might produce observed signatures in future high energy
collider experiments.

Among various kinds of technicolor theories, the topcolor
scenario\cite{topcolor} is attractive because it can explain the
large top quark mass and provides a possible EWSB mechanism. The
topcolor-assisted technicolor (TC2) model\cite{tc2-rev} is one of
the phenomenologically viable models, which has all essential
features of the topcolor scenario. This model predicts three CP-odd
top-pions $\pi_t^0,~\pi_t^\pm$ and one CP-even top-higgs $h_t^0$
with large couplings to the third family, which may make these new
scalar particles have a distinct experimental
signature\cite{works-tc2}. Thus, discovery of a doubly scalar
particles in future high energy colliders would be a definite signal
of new physics beyond the SM, which would help us to understand the
scalar sector and more importantly what lies beyond the SM.

The CERN Large Hadron Collider (LHC) has already started its
operation, and it will have considerably capability to discover and
measure almost all the quantum properties of a standard model (SM)
higgs boson of any mass \cite{mh-lhc}. However, from the theoretical
view point, it would be expected that the SM is replaced by a more
fundamental theory at the TeV scale. If hadron colliders find
evidence for a new scalar state, it may not necessarily be the SM
higgs boson. Many alternative new physics theories, such as
supersymmetry, technicolor, and little Higgs, predict the existence
of new scalars or pseudo-scalar particles. These new particles may
have so large cross sections and branching fractions  as to be
observable at the high energy colliders. Thus, studying the
production and decays of the new scalars at hadron colliders, the
future international lepton collider and the $\gamma\gamma$ collider
will be of special interest.

On the other side, at the tree-level or one-loop level, the scalar
pair productions of the new particles predicted in the new physics
model may have very large production rates \cite{pair-production},
so it may be interesting to consider the pair production of the new
scalars and analysis the observable possibility in TC2 model. We
hope that $SS'$ ($S,~S'$ denotes any one of the new scalars, i.e,
top-pions $\pi_t^0,~\pi_t^\pm$ and top-higgs $h_t^0$) productions
can be carried out at the LHC, the future international linear
collider (ILC) and the photon linear collider (PLC) to test the
topcolor scenario of the TC2 model.

At the LHC, the measurement of the pair production would be
interesting. If we can get larger cross sections surpassing the SM
prediction, it would provide a powerful proof to probe the new
physics model.
At the ILC,  A PGB pair can be produced in the process $e^+e^-\to
SS'$, in which the $SS'Z(\gamma)$ coupling can be probed via the new
couplings $Z(\gamma)\pi_t^+\pi_t^-$ and $Z\pi_t^0 h_t^0$. On the
other hand, the PLC option may also be useful to explore the new
physics coupling concerning the PGBs. At the second stage of the ILC
($\sqrt{s}=1.5$ TeV),  the signal should increase with the
increasing center-of-mass.

In this paper, we study how the technicolor models affect the scalar
pair production processes $gg\to SS'$, $e^+e^-\to SS'$, $q\bar q \to
SS'$ and $\gamma\gamma\to SS'$, via the new couplings in the TC2
model. Cross sections for these bosons pair production processes are
evaluated, and can be significantly large considering the small SM
backgrounds. By measuring these double scalar bosons production
processes at different collider experiments, we would be able to
probe properties of new physics particles, which helps identify the
new physics model.

In Sec.~II, the technicolor models relative to our calculations are
reviewed, and the effects of the new couplings in the scalar boson
pair production processes $gg\to SS'$ and $q\bar q\to
SS'$($q=u,d,c,s,b$ quarks) at LHC, $e^+e^-\to SS'$ at the ILC, and
$\gamma\gamma\to SS'$ at the photon collider options, discussed too.
Sec.~III shows the the numerical results for every processes,
respectively and analysis the SM backgrounds and the detectable
probability for every final state at the different colliders.
Summary and discussions are given in Sec.~IV.

\section{TC2 model and the relevant couplings}
To solve the phenomenological difficulties of traditional
technicolor(TC) theory, TC2 theory\cite{tc2-rev} was proposed by
combing TC interactions with the topcolor interactions for the third
 generation at the scale of about 1 TeV.


The TC2 theory introduces two strongly interacting sectors, with one
sector (topcolor interaction) generating the large top quark mass
and partially contributing to EWSB while the other sector
(technicolor interaction) responsible for the bulk of EWSB and the
generation of light fermion masses.  At the EWSB scale, it predicts
the existence of two groups of composite scalars from topcolor and
technicolor condensations, respectively
\cite{tc2-rev,top-condensation}. In the linear realization, the
scalars of our interest can be arranged into two $SU(2)$ doublets,
namely  $\Phi_{top}$ and $\Phi_{TC}$
\cite{top-condensation,2hd,0110218-Rainwater}, which are analogous
to the Higgs fields in a special two-Higgs-doublet model
\cite{special-thdm}. The doublet $\Phi_{top}$ from topcolor
condensation couples only to the third-generation quarks. Its main
task is to generate the large top quark mass. It can also generate a
sound part of bottom quark mass indirectly via instanton
effect\cite{tc2-rev}. Since a small value of the top-pion decay
constant $F_t $ (the vacuum expectation value (VEV) of the doublet
$\Phi_{top}$) is theoretically favored (see below), this doublet
must couple strongly to top quark in order to generate the expected
top quark mass. The other doublet $\Phi_{TC}$, which is technicolor
condensate,  is mainly responsible for EWSB and light fermion
masses. It also contributes a small portion to the third-generation
quark masses. Because its VEV $V_{TC}$ is generally comparable with
$V_W$, its Yukawa couplings with all fermions are small. The Yukawa
term in the low-energy effective Lagrangian can be written as
\cite{0110218-Rainwater}
\begin{eqnarray}
{\cal L}_Y=  - \left ( \sum_{i,j=1}^3 \lambda_{i j}^U \bar{Q}_{L i}
\Phi_{TC} U_{R j} + \sum_{i,j=1}^3 \lambda_{i j}^D \bar{Q}_{L i}
\tilde{\Phi}_{TC} D_{R j} + Y_t \bar{\Psi}_L \Phi_{top} t_R + h.c.
\right ) + \cdots
\end{eqnarray}
where  $Q_{L i}$ denotes the left-handed quark doublet, $U_{R j}$
and $ D_{R j}$ are right-handed quarks, $\Psi_L $ is the left-handed
top-bottom doublet, $\tilde{\Phi}_{TC} $ is the conjugate of
$\Phi_{TC}$, and $\lambda_{i j}^{U, D}$ and $Y_t $ are Yukawa
coupling constants satisfying $\lambda_{i j}^{U, D}\ll Y_t$. The two
$SU(2)$ doublets take the form
\begin{eqnarray}
\Phi_{TC}& =&\left ( \begin{array}{c} V_{TC} + ( H_{TC}^0 + i \Pi_{TC}^0 )/\sqrt{2}  \\
           \Pi_{TC}^- \end{array} \right ) , \label{phi1}\\
\Phi_{top}& =& \left (\begin{array}{c} F_t + ( H_{top}^0 + i \Pi_{top}^0 )/\sqrt{2}  \\
           \Pi_{top}^- \end{array} \label{phi2}\right ) .
\end{eqnarray}
We can rotate the two doublets into $\Phi_{1,2}$ such that
$<\Phi_1>=\sqrt{V_{TC}^2 + F_t^2}=V_W$ and $<\Phi_2> =0$
\begin{eqnarray}
\Phi_1 & = & (\cos \beta \Phi_{TC} + \sin \beta \Phi_{top} )=
            \left ( \begin{array}{c} v_{w} + ( H_1^0+ i G^0 )/\sqrt{2}  \\
             G^- \end{array} \right ) , \\
\Phi_2 & = & (- \sin \beta \Phi_{TC} + \cos \beta \Phi_{top} )
            =\left (
            \begin{array}{c} ( H_2^0 + i A^0 )/\sqrt{2} , \\
            H^- \end{array} \right ) ,
\end{eqnarray}
where $\tan \beta =F_t/V_{TC}$. Then the Lagrangian can be rewritten
as
\begin{eqnarray}
{\cal L}_Y&=& - \left ( \sum_{i,j=1}^3 \lambda_{i j}^{\prime U}
\bar{Q}_{L i} \Phi_{1} U_{R j} + \sum_{i,j=1}^3 \lambda_{i j}^D
\frac{\sqrt{V_W^2-F_t^2}}{V_W} \bar{Q}_{L i} \tilde{\Phi}_{1} D_{R
j}   - \sum_{i,j=1}^3 \lambda_{i j}^D \frac{F_t}{V_W} \bar{Q}_{L i}
\tilde{\Phi}_{2} D_{R j} \right. \nonumber   \\  & & \left.-
\sum_{i,j=1}^3 \lambda_{i j}^U \frac{F_t}{V_W} \bar{Q}_{L i} \Phi_2
U_{R j}+ Y_t \frac{\sqrt{V_W^2-F_t^2}}{V_W} \bar{\Psi}_L \Phi_{2}
t_R + h.c. \right ) + \cdots \label{Lagrangian}
\end{eqnarray}
where $\lambda_{i j}^{\prime U} =\lambda_{i j}^U \cos \beta + Y_t
\sin \beta \delta_{i 3} \delta_{j 3} $. In this new basis, $G^\pm$
and $G^0$ are Goldstone bosons while the pseudoscalar $A^0$, the
charged scalar $H^\pm$ and the CP-even scalars $H_{1,2}^0$ are
physical PGBs.
It is obvious that $H_1^0$ plays the role of the "standard" Higgs
boson with flavor diagonal couplings and  $H_2^0$ decouples from the
SM vector bosons but has strong coupling only with top quark. In our
following analysis, we will adopt the same notations as in the
literature, i.e.,  using top-Higgs $h_t^0$, top-pions
$\pi_t^{0,\pm}$ to denote $H_2^0$, $A^0$ and $H^\pm$, respectively.

In Eq.(\ref{Lagrangian}), the rotation of quarks into their mass
eigenstates will induce FCNC Yukawa interactions from the $\Phi_2$
couplings \footnote{ Just like the Higgs field in the SM, $\Phi_1$
terms give no FCNC couplings since they are diagonalized
simultaneously with the fermion mass matrices.}. Since $\lambda_{i
j}^{U, D}\ll Y_t $, the FCNC couplings from $\lambda_{i j}^U$ and
$\lambda_{i j}^D$ can be safely neglected.  Because $Y_t
=(1-\epsilon) m_t/F_t$ ($\epsilon $ denoting the fraction of
technicolor contribution to the top quark mass) is quite large
(about $2 \sim 3 $) and the mixing between $c_R$ and $t_R$ can be
naturally as large as 30\% \cite{FCNH}, the FCNC coupling from the
$Y_t$ term may be sizable and thus may have significant
phenomenological consequence. The FCNC couplings from this term are
given by
\begin{eqnarray}
{\cal{L}}_{FCNC}& = &\frac{(1 - \epsilon ) m_{t}}{\sqrt{2}F_{t}}
     \frac{\sqrt{v_{w}^{2}-F_{t}^{2}}} {v_{w}} \left (
            i K_{UL}^{tt*}K_{UR}^{tt } \bar{t}_L t_{R} \pi_t^0
           + \sqrt{2}K_{UR}^{tt *} K_{DL}^{bb}\bar{t}_R b_{L} \pi_t^-
           + i K_{UL}^{tt *} K_{UR}^{tc} \bar{t}_L c_{R} \pi_t^0  \right . \nonumber \\
& & \left. + \sqrt{2} K_{UR}^{tc *} K_{DL}^{bb} \bar{c}_R b_{L}
\pi_t^-
           + K_{UL}^{tt*} K_{UR}^{tt } \bar{t}_L t_{R} h_t^0
           + K_{UL}^{tt *} K_{UR}^{tc} \bar{t}_L c_{R} h_t^0 + h.c.  \right ) ,
\label{FCNH}
\end{eqnarray}
where $K_{UL}$, $K_{DL}$ and $K_{UR}$ are the rotation matrices that
transform the weak eigenstates of left-handed up-type, down-type and
right-handed up-type quarks to their mass eigenstates, respectively.
According to the analysis of \cite{FCNH}, their favored values are
given by
\begin{equation}
K_{UL}^{tt} \simeq K_{DL}^{bb} \simeq 1, \hspace{5mm}
K_{UR}^{tt}\simeq \frac{m_t^\prime}{m_t} = 1-\epsilon, \hspace{5mm}
K_{UR}^{tc}\leq \sqrt{1-(K_{UR}^{tt})^2}
=\sqrt{2\epsilon-\epsilon^{2}}, \label{FCSI}
\end{equation}
with $m_t^\prime$ denoting the topcolor contribution to the top
quark mass. In Eq.(\ref{FCNH}) we neglected the mixing between up
quark and top quark.

Using the same scalar SU(2) doublets in Eq.
(\ref{phi1}),(\ref{phi2}), the kinetic term is
\begin{equation}
{\cal L}_{kin} \, = \,
  \biggl( D_\mu \Phi_{TC}  \biggr)^\dagger \biggl( D^\mu \Phi_{TC}  \biggr)
+ \biggl( D_\mu \Phi_{top} \biggr)^\dagger \biggl( D^\mu \Phi_{top}
\biggr), \label{eq:L-kin}
\end{equation}
The covariant derivative is
\begin{equation}
D_\mu \, = \, \partial_\mu + i {g' Y\over 2} B_\mu
   + i {g \over 2} \tau_i W^i_\mu \, .
\label{eq:Dmu}
\end{equation}
The hypercharge of the doublets is $Y = -1$. We make the following
redefinition of fields:
\begin{eqnarray}
W^\pm_\mu = {1\over\sqrt{2}}(W^1_\mu \mp iW^2_\mu), \\
W^3_\mu =   Z_\mu \cos\theta + A_\mu \sin\theta,      \\
B_\mu   = - Z_\mu \sin\theta + A_\mu \cos\theta. \label{eq:V-fields}
\end{eqnarray}
After replacement of the physical vector boson fields, the $D_\mu
\Phi_i$ term for each doublet will be of the form
\begin{eqnarray}
D_\mu \Phi_i \, = & \left(\begin{array}{c}
{1\over\sqrt{2}}(\partial_\mu H_i + i\partial_\mu\Pi^0_i) \\
i\partial_\mu\Pi^-_i
\end{array}\right)
& \; + \; {i g_Z\over 2} \, Z_\mu \binom{{1\over\sqrt{2}}(v_i + H_i
+ i\Pi^0_i)}{-i(1-2\sin^2\theta_W)\Pi^-_i}
\nonumber\\
& \; + \; e A_\mu \left(\begin{array}{c}
0 \\
\Pi^-_i
\end{array}\right)
& \; + \; {i g\over 2} \binom{i\sqrt{2} \, W^+_\mu\Pi^-_i}{W^-_\mu
(v_i + H_i + i\Pi^0_i)}. \label{eq:DmuHi}
\end{eqnarray}
where $g_Z = g/\cos\theta_W$ and $e = g\sin\theta_W$. After
expanding the terms in Eq.~(\ref{eq:L-kin}), we form orthogonal
linear combinations of the fields $\Pi^{0,\pm}_i$,
\begin{eqnarray}
G^{0,\pm}   \; & = & \; {F_t \Pi^{0,\pm}_{top} + V_{TC}
\Pi^{0,\pm}_{TC} \over V_W}
\; \; \; {\rm (Goldstone \; bosons)}, \\
\pi_t^{0,\pm} \; & = & \; {V_{TC}   \Pi^{0,\pm}_{top} - F_t
\Pi^{0,\pm}_{TC} \over V_W} \; \; \; {\rm (physical \; top-pions)}.
\label{eq:pions}
\end{eqnarray}
%

After rearrangement the Feynman rules can simply be read off, however,
Table~\ref{tbl:coups1} lists only the 3-point gauge couplings for the
physical fields relative to our calculation.
\begin{table}[htb]
\begin{tabular}{|c|l||c|l|}
\hline

$Z^\mu h_t^0\pi_t^0$ & $-{ig_Z\over 2 } {V_{TC}   \over V_W} \,
(p^h_\mu - p^0_\mu)$ & $A^\mu \pi_t^- \pi_t^+$   & $e \, (p^-_\mu -
p^+_\mu)$
 \\
$Z^\mu \pi_t^- \pi_t^+$   & ${g_Z} \, (1-2\sin^2\theta_W) \,
(p^-_\mu - p^+_\mu)$ & $W^{-\mu} \pi_t^0   \pi_t^+$ & $-{g\over 2}
 (p^0_\mu - p^+_\mu)$
                   \\
$W^{-\mu} h_t^0\pi_t^+$ & $-{ig\over 2} {V_{TC}   \over V_W} \,
(p^h_\mu - p^+_\mu)$ & $W^{+\mu} \pi_t^-  h_t^0$ & ${ig\over 2}
{V_{TC}   \over V_W} \, (p^-_\mu - p^h_\mu)$  \\\hline
\end{tabular}
\vspace{2mm} \caption[]{\label{tbl:coups1}
 3-point TC2 gauge couplings for the physical fields;
     All bosons (charge and momentum) flow out.}
\end{table}

Now we recapitulate the theoretical and experimental constraints on
the relevant parameters.
\begin{itemize}
\item[{\rm (1)}] About the $\epsilon$ parameter. In the TC2 model,
$\epsilon $ parameterizes the portion of the extended-technicolor
(ETC) contribution to the top quark mass. The bare value of
$\epsilon $ is generated at the ETC scale, and subject to very large
radiative enhancement from the topcolor and $U(1)_{Y_1}$ by a factor
of order $10$ when evolving down to the weak scale \cite{tc2-rev}.
This $\epsilon$ can induce a nonzero top-pion mass (proportional to
$\sqrt{\epsilon} $) \cite{Hill} and thus ameliorate the problem of
having dangerously light scalars. Numerical analysis shows that,
with reasonable choice of other input parameters, $\epsilon$ of
order $10^{-2} \sim 10^{-1}$ may induce top-pions as massive as the
top quark \cite{tc2-rev}. Indirect phenomenological constraints on
$\epsilon $ come from low energy flavor-changing processes such as
$b \to s \gamma$ \cite{b-sgamma}. However, these constraints are
very weak. From the theoretical point of view, $\epsilon $ with
value from $0.01$ to $0.1$ is favored. Since a large $\epsilon$ can
slightly suppress the FCNC Yukawa couplings, we fix conservatively
$\epsilon =0.1$ throughout this paper.

\item[{\rm (2)}] The parameter $K_{UR}^{tc}$ is upper bounded by the unitary relation
$K_{UR}^{tc} \leq \sqrt{1-(K_{UR}^{tt})^ 2}=\sqrt{2\epsilon
-\epsilon^2}$. For a $\epsilon $ value smaller than $0.1 $, this
corresponds to $ K_{UR}^{tc} < 0.43$. In our analysis, we will treat
$K_{UR}^{tc}$ as a free parameter.

\item[{\rm (3)}] About the top-pion decay constant $F_t$,  the Pagels-Stokar formula \cite{Pagels}
gives an expression in terms of the number of quark color $N_c$, the
top quark mass, and the scale $\Lambda $ at which the condensation
occurs:
\begin{eqnarray}
F_t^2= \frac{N_c}{16 \pi^2} m_t^2 \ln{\frac{\Lambda^2}{m_t^2}}.
\label{ft}
\end{eqnarray}
From this formula, one can infer that, if $t\bar{t} $ condensation
is fully responsible for EWSB, i.e. $F_t \simeq V_W \equiv
v/\sqrt{2} = 174$ GeV, then $\Lambda $ is about $10^{13} \sim
10^{14}$ GeV. Such a large value is less attractive since by the
original idea of technicolor \cite{technicolor}, one expects new
physics scale should not be far higher than the weak scale. On the
other hand, if one believes that new physics exists at TeV scale,
i.e. $\Lambda \sim 1$ TeV, then $F_t \sim 50$ GeV, which means that
$t \bar{t} $ condensation alone cannot be wholly responsible for
EWSB and to break electroweak symmetry needs the joint effort of
topcolor and other interactions like technicolor. By the way,
Eq.(\ref{ft}) should be understood as only a rough guide, and $F_t$
may in fact be somewhat lower or higher, say in the range $40 \sim
70$ GeV. Allowing $F_t $ to vary over this range does not
qualitatively change our conclusion, and, therefore, we use the
value $F_t =50$ GeV for illustration in our numerical analysis.

\item[{\rm (4)}] About the mass bounds for top-pions and top-Higgs.
On the theoretical side, some estimates have been done. The mass
splitting between the neutral top-pion and the charged top-pion
should be small since it comes only from the electroweak
interactions \cite{mass-pion}. Ref.\cite{tc2-rev} has estimated the
mass of top-pions using quark loop approximation and showed that
$m_{\pi}$ is allowed to be a few hundred GeV in a reasonable
parameter space. Like Eq.(\ref{ft}), such estimations can only be
regarded as a rough guide and the precise values of top-pion masses
can be determined only by future experiments. The mass of the
top-Higgs $h_t^0$ can be estimated in the Nambu-Jona-Lasinio (NJL)
model in the large $N_{c}$ approximation and is found to be about
$2m_{t}$ \cite{top-condensation,top-Higgs}. This estimation is also
rather crude and the mass below the $\overline{t}t$ threshold is
quite possible in a variety of scenarios \cite{y15}. On the
experimental side,  current experiments have restricted the mass of
the charged top-pion. For example, the absence of $t \to \pi_t^+b$
implies that $m_{\pi_t^+} > 165$ GeV \cite{t-bpion} and $R_b$
analysis yields $m_{\pi_t^+}> 220$ GeV \cite{burdman,kuang}. For the
neutral top-pion and top-Higgs, the experimental restrictions on
them are rather weak. (Of course, considering theoretically that the
mass splitting between the neutral and charged top-pions is small,
the $R_b$ bound on the charged top-pion mass should be applicable to
the neutral top-pion masses.)  The current bound on techni-pions
\cite{datagroup} does not apply here since the properties of
top-pion are quite different from those of techni-pions. The direct
search for the neutral top-pion (top-Higgs) via $ p p ({\rm
or}~p\bar p) \to t \bar{t} \pi_t^0 (h_t^0)$ with $\pi_t^0 (h_t^0)
\to b \bar{b} $ was proven to be hopeless at Tevatron for the
top-pion (top-Higgs) heavier than $135 $ GeV
\cite{0110218-Rainwater}. The single production of $\pi_t^0 $
($h_t^0$ ) at Tevatron with $\pi_t^0 $ ($h_t^0$) mainly decaying to
$t \bar{c} $ may shed some light on detecting top-pion
(top-Higgs)\cite{top-Higgs}, but the potential for the detection is
limited by the value of $K_{UR}^{tc}$ and the detailed background
analysis is absent now. Moreover, these mass bounds will be greatly
tightened  at the running and the incoming LHC
\cite{cao1,FCNH,0110218-Rainwater}, and Ref.\cite{1108.4000} has
limited the top-pion mass larger than 300 GeV. Combining the above
theoretical and experimental bounds,  we in our discussion will
assume
\begin{equation}
m_{h_t^0} > 300 ~{\rm GeV} \hspace{5mm}
m_{\pi_{t}^{0}}=m_{\pi_{t}^+}\equiv m_{\pi_t} > 220 ~{\rm GeV} .
\end{equation}
\end{itemize}
We, however, in the following calculations will assume the top-Higgs
mass equal to that of the top-pion and see the behavior in the
assumption.

\section{The PGB pair productions at colliders}
In this section, we discuss PGB pair production processes $gg\to
SS'$, $q\bar q SS'$($q=~u,~d,~s,~c,~b$), $e^+e^-\to SS'$ and
$\gamma\gamma\to SS'$ in TC2 model. In these processes, some
couplings such as $\pi_t^\pm \bar b c$ and $Z\pi_t^+\pi_t^-$,
$Z\pi_t^0 h_t^0$ etc.,  contain the model-dependent parameters so
that they can be used to probe the new physics theory at future
collider experiments.

At the LHC, the cross sections of the PGB pair production comes
mainly from the gluon fusion and quark pair annihilation processes
$gg\to \pi_t^+\pi_t^-$, $gg\to
\pi_t^0h_t^0,~\pi_t^0\pi_t^0,~h_t^0h_t^0$,  $q\bar q\to
\pi_t^+\pi_t^-$, $q\bar q \to
\pi_t^0h_t^0,~\pi_t^0\pi_t^0,~h_t^0h_t^0$($q=u,d,s,c,b$), $u\bar
d\to \pi_t^+\pi_t^0(h_t^0)$.

Note that for the neutral final states, there could be $\pi_t^0
\pi_t^0 $,  $\pi_t^0 h_t^0 $ and $h_t^0h_t^0$, the cross sections of
which, however, are not the same even if we take $m_{h_t}$ equal to $m_{\pi}$
with the same values of the other parameters, and in the following,
we will discuss them one by one.  It is relevant to calculate separately
the cross sections for the $\pi_t^0\pi_t^0 $,  $\pi_t^0 h_t^0 $ and $h_t^0h_t^0$,
final states because, for equal masses
of $\pi^0_t$ and $h^0_t$, all three of these final states will contribute to the
experimental signal, and so it is important to know the cross sections
for all three of them.

At the LHC, the parton level cross sections 
 are calculated at the leading order as
\begin{equation}
\hat{\sigma}(\hat s)
=\int_{\hat{t}_{min}}^{\hat{t}_{max}}\frac{1}{16\pi \hat{s}^2}
\overline{\Sigma}|M_{ren}|^{2}d\hat{t}\,,
\end{equation}
with
\begin{eqnarray}
\hat{t}_{max,min} =\frac{1}{2}\left\{ m_{p_1}^2 +m_{p_2}^2 -\hat{s}
\pm \sqrt{[\hat{s} -(m_{p_1}+m_{p_2})^2][\hat{s} -(m_{p_1}
-m_{p_2})^2]} \right\},
\end{eqnarray}
where $p_1$ and $p_2$ are the first and the second initial particles
in the parton level, respectively. For our case, they could be gluon
$g$ and quarks $u$, $d$, $c$, $s$, $b$ etc.

 The total hadronic cross section for $pp \rightarrow SS'+X$ can be obtained by folding the
subprocess cross section $\hat{\sigma}$ with the parton luminosity
\begin{equation}
\sigma(s)=\int_{\tau_0}^1 \!d\tau\, \frac{dL}{d\tau}\, \hat\sigma
(\hat s=s\tau) ,
\end{equation}
where $\tau_0=(m_{p_1}+m_{p_2})^2/s$, and $s$ is the $p p$
center-of-mass energy squared. $dL/d\tau$ is the parton luminosity
given by
\begin{equation}
\frac{dL}{d\tau}=\int^1_{\tau} \frac{dx}{x}[f^p_{p_1}(x,Q)
f^{p}_{p_2}(\tau/x,Q)+(p_1\leftrightarrow p_2)],\label{dis-pp}
\end{equation}
where $f^p_{p_1}$ and $f^p_{p_2}$ are the parton $p_1$ and $p_2$
distribution functions in a proton, respectively.
 In our numerical
calculation, the CTEQ6L parton distribution function is
used~\cite{Ref:cteq6} and take factorization scale $Q$ and the
renormalization scale $\mu_F$ as $Q=\mu_F = 2 m_{\pi_1}$. The loop
integrals are evaluated by the LoopTools
package~\cite{Ref:LoopTools}.

At an electron-positron linear collider, the PGB pair production can
be realized by $e^+e^-\to \pi_t^+\pi_t^-$, $e^+e^-\to \pi_t^0
h_t^0$. The $e^+e^-\to \pi_t^+\pi_t^- ,\pi_t^0 h_t^0$ process may be
promising channels at the ILC for light PGBs because of the simple
kinematical structure. Since relatively larger collision energy is
required for two top-pion final states of $\pi_t^+\pi_t^-$ and
$\pi_t^0 h_t^0$, the $s$-channel nature of the process may decrease
the cross section. On the other hand, if we have large enough
energy, one can control the collision energy to obtain the maximal
production rate.

At a high energy lepton collider, the hard photons can be obtained
from the Compton back scattering method~\cite{photon collider}. By
using hard photons, PGB pairs can be produced in
$\gamma\gamma\to\pi_t^+\pi_t^-$ and $\gamma\gamma\to\pi_t^0h_t^0$,
$\pi_t^0\pi_t^0$ and $h_t^0h_t^0$ processes, the feynman diagrams of
which are shown in FIG.~\ref{rrpp-fey}.

Since the photon beams in $\gamma\gamma$ collision are generated by
the backward Compton scattering of the incident electron- and the
laser-beam, the events number is obtained by convoluting the cross
section of $\gamma\gamma$ collision with the photon beam luminosity
distribution:
\begin{eqnarray}
N_{\gamma \gamma \to \ell_i \bar\ell_j}&=&\int
d\sqrt{s_{\gamma\gamma}}
  \frac{d\cal L_{\gamma\gamma}}{d\sqrt{s_{\gamma\gamma}}}
  \hat{\sigma}_{\gamma \gamma \to \ell_i \bar\ell_j}(s_{\gamma\gamma})
  \equiv{\cal L}_{e^{+}e^{-}}\sigma_{\gamma \gamma \to \ell_{i} \bar\ell_{j}}(s)
\end{eqnarray}
where $d{\cal L}_{\gamma\gamma}$/$d\sqrt{s}_{\gamma\gamma}$ is the
photon-beam luminosity distribution and $\sigma_{\gamma \gamma \to
\ell_i \bar\ell_j}(s)$ ( $s$ is the squared center-of-mass energy of
$e^{+}e^{-}$ collision) is defined as the effective cross section of
$\gamma \gamma \to \ell_{i} \bar\ell_{j}$. In the optimum case, it
can be written as \cite{photon collider}
\begin{eqnarray}
\sigma_{\gamma \gamma \to \ell_i \bar\ell_j}(s)&=&
  \int_{\sqrt{a}}^{x_{max}}2zdz\hat{\sigma}_{\gamma \gamma \to \ell_{i} \bar\ell_{j}}
  (s_{\gamma\gamma}=z^2s) \int_{z^{2/x_{max}}}^{x_{max}}\frac{dx}{x}
 F_{\gamma/e}(x)F_{\gamma/e}(\frac{z^{2}}{x})
\end{eqnarray}
where $F_{\gamma/e}$ denotes the energy spectrum of the
back-scattered photon for the unpolarized initial electron and laser
photon beams given by
\begin{eqnarray}
F_{\gamma/e}(x)&=&\frac{1}{D(\xi)}\left[1-x+\frac{1}{1-x}-\frac{4x}{\xi(1-x)}
  +\frac{4x^{2}}{\xi^{2}(1-x)^{2}}\right]
\end{eqnarray}
with
\begin{eqnarray}
D(\xi)&=&(1-\frac{4}{\xi}-\frac{8}{\xi^{2}})\ln(1+\xi)
  +\frac{1}{2}+\frac{8}{\xi}-\frac{1}{2(1+\xi)^{2}}.
\end{eqnarray}
Here $\xi=4E_{e}E_{0}/m_{e}^{2}$ ($E_{e}$ is the incident electron
energy and $E_{0}$ is the initial laser photon energy) and
$x=E/E_{E}$ with $E$ being the energy of the scattered photon moving
along the initial electron direction. The definitions of parameters
$\xi$, $D(\xi)$ and $x_{max}$ can be found in Ref.\cite{photon
collider}. In our numerical calculation, we choose $\xi=4.8$,
$D(\xi)=1.83$ and $x_{max}=0.83$.


\section{The PGB pair productions in $pp$, $e^+e^-$
and $\gamma\gamma$ collisions}
 In this section, we study cross sections for the
double PGB production processes $gg\to \pi_t^+\pi_t^- $,  $\pi_t^0
h_t^0$, $\pi_t^0\pi_t^0$, $h_t^0h_t^0$, $q\bar q \to \pi_t^+\pi_t^-,
\pi_t^0 h_t^0$, $\pi_t^0\pi_t^0$, $h_t^0h_t^0$, $e^+e^-\to
\pi_t^+\pi_t^-, \pi_t^0 h_t^0$, and $\gamma\gamma\to
\pi_t^+\pi_t^-$, $\pi_t^0 h_t^0$, $\pi_t^0\pi_t^0$, $h_t^0h_t^0$.
Since the signals of these processes as well as their corresponding
backgrounds are not the same, we will analysis these processes
separately. Throughout this paper, we take $m_t =173 $ GeV
\cite{topmass}, $m_W=80.38 $ GeV, $m_Z=91.19 $ GeV \cite{datagroup},
$\alpha_s(m_Z) = 0.118 $ and neglect bottom quark mass as well as
charm quark mass.

As for the TC2 parameters, we will consider the masses of the
scalars equal to each other, i.e. the masses of the top-pions,
neutral and charged, denoted as $m_\pi$ when not considering the
difference between them. Considering the discussion in the previous
section, we will take $m_\pi$ and $K_{UR}^{tc}$ as the free
parameters and assume $m_\pi$ are in the range $200 - 600$ GeV,
$K_{UR}^{tc} = 0.1 - 0.4$.

\subsection{At The LHC}
The parton processes $gg\to \pi_t^+\pi_t^- $, $\pi_t^0 h_t^0$,
$\pi_t^0\pi_t^0$, $h_t^0h_t^0$, $q\bar q \to \pi_t^+\pi_t^-, \pi_t^0
h_t^0, \pi_t^+ \pi_t^0, h_t^0 h_t^0$ can be produced at the LHC,
with the feynman diagrams shown in Fig.\ref{ggpp-fey} and
Fig.\ref{qqpp-fey}. To relatively know them, we here, firstly,
discuss the contributions from every single parton channel though ,
actually, we can not distinguish the initial states, i.e, we will
firstly discuss  the $gg$ fusion and the $q\bar q$ annihilation
processes, respectively, and then sum them all together to see the
total contributions.

By contrast to the lepton collider, the situation would be
deteriorated at the hadron colliders such as LHC, however, on other
aspect, the production probability of the new physics particles at
LHC may be much larger, so that the disadvantage caused by
background contamination may be compensated, which is proven in the
following discussions.

\subsubsection{$gg\to \pi_t^+\pi_t^- $ and $gg\to \pi_t^0 h_t^0 $, $\pi_t^0 \pi_t^0 $, $h_t^0 h_t^0 $}
\begin{figure}[hbt]
\begin{center}
\epsfig{file=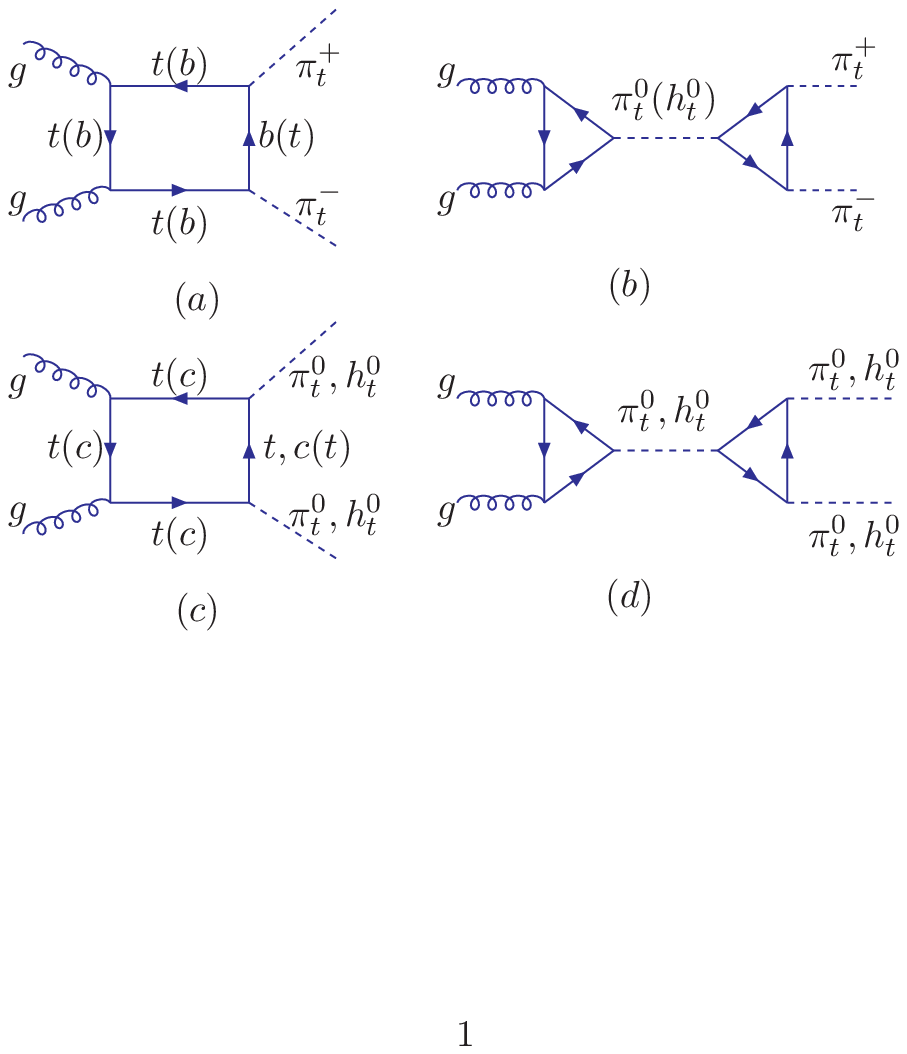,width=12cm}
\caption{Feynman diagrams for the PGB pair production at the LHC via
gluon fusion parton level processes in the TC2 model. Those obtained
by exchanging the two external gluon lines are not displayed here.}
\label{ggpp-fey}
\end{center}
\end{figure}

Due to the interactions in Eq.(\ref{FCNH}), the PGB pair production
processes can proceed through various parton processes at the LHC,
as shown in Fig.\ref{ggpp-fey}, in which those obtained by
exchanging the two external gluon lines are not displayed here.

Note that the processes consist of the box diagrams and the
trilinear scalar coupling \cite{0510201}, just shown as
Fig.\ref{ggpp-fey}(a)(c) and (b)(d). The box contribution of the
cross sections, however, is dominant since, firstly, in the
s-channel contribution, the center of mass depress the production
rate. Secondly, note that the $\pi_t^+ t \bar b$ coupling strength,
$Y\sim \frac{m_t}{F_t}\frac{\sqrt{v_{W}^{2}-F_{t}^{2}}} {v_{W}} \sim
3 $, so we can imagine that one expects they may induce larger
contributions to the relevant processes. Finally, the trilinear
scalar coupling, i.e., the s-channel contribution is a two-loop
diagram, as was expected, the contribution should be smaller than
that of the one-loop contribution, i.e, the box diagram. The two
contributions, we have calculated, are very small, less than $1$ fb,
and the interference contributions are small too, so we will not
discuss them in the followings.

\def\figsubcap#1{\par\noindent\centering\footnotesize(#1)}
\begin{figure}[bht]%
\begin{center}
\hspace{-0.25cm}
\parbox{8.05cm}{\epsfig{figure=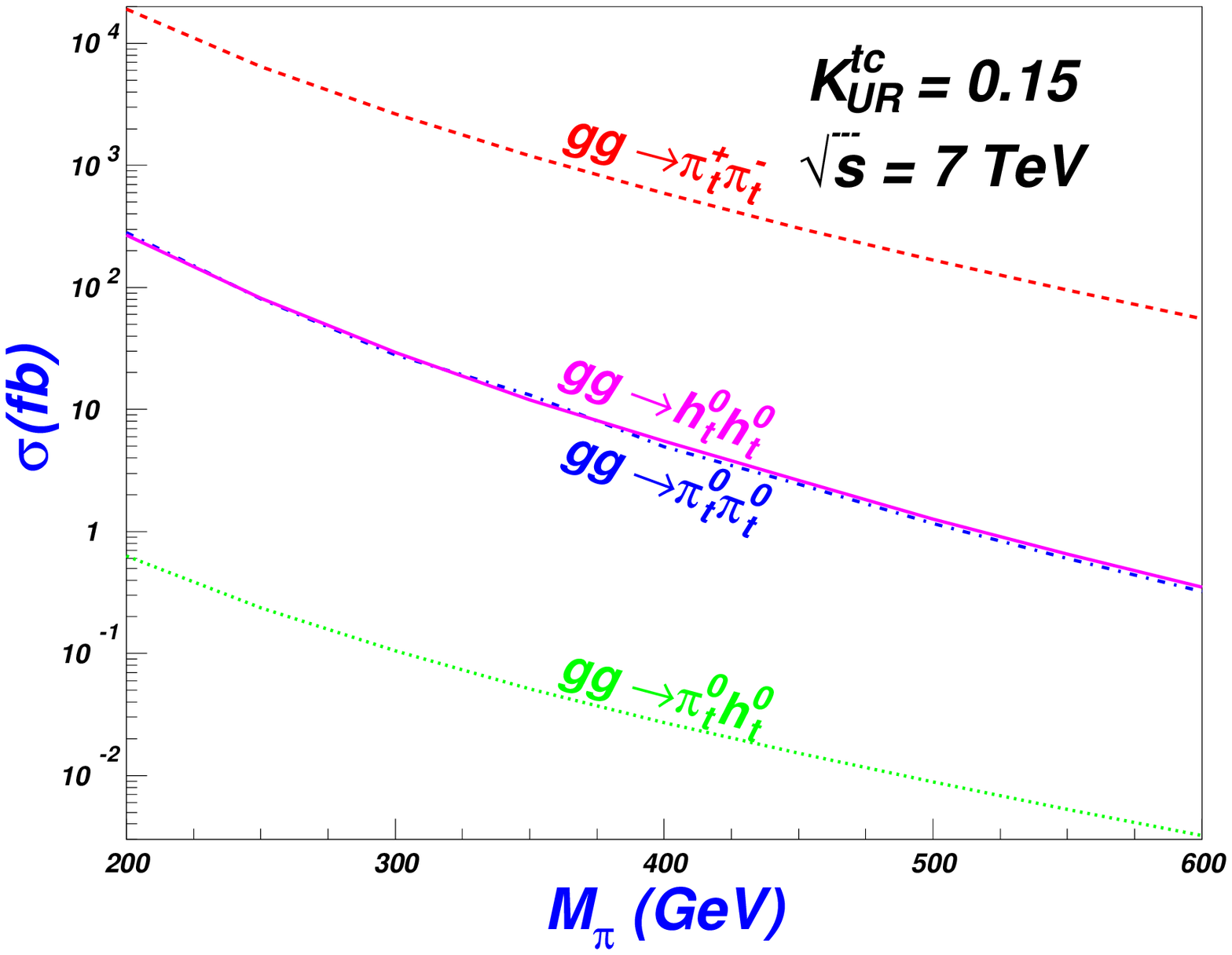,width=8cm} }
\parbox{8.05cm}{\epsfig{figure=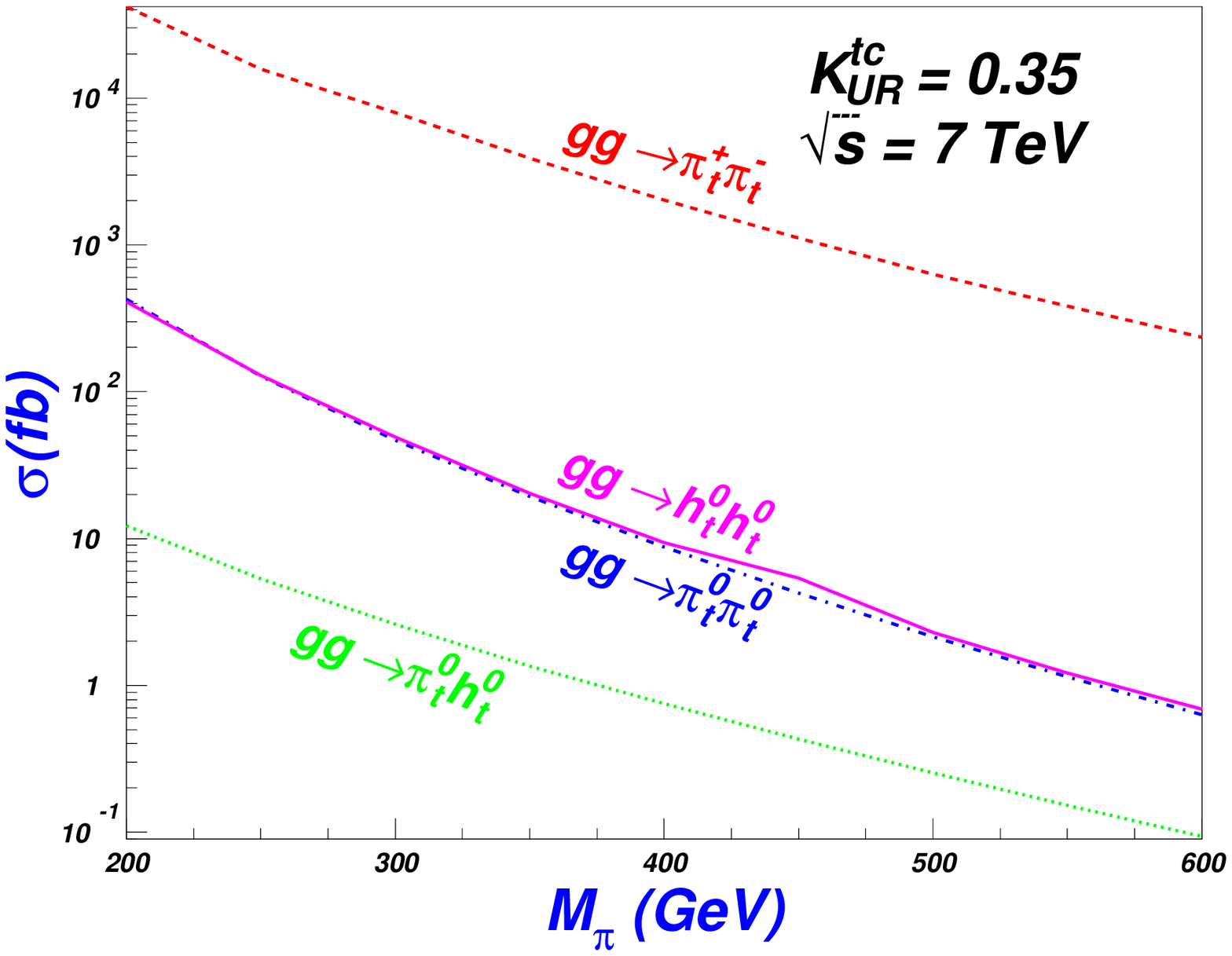,width=8cm} }
\caption{ The cross section $\sigma$ of the processes $gg\to SS' $
as a function of the top-pion mass $m_{\pi_t}$ with
$K_{UR}^{tc}=0.15$ and $K_{UR}^{tc}=0.35$ and $\sqrt{s}=7$ TeV.
 \label{ggpp1}  }
\end{center}
\end{figure}

\def\figsubcap#1{\par\noindent\centering\footnotesize(#1)}
\begin{figure}[bht]%
\begin{center}
\hspace{-0.25cm}
 \parbox{8.05cm}{\epsfig{figure=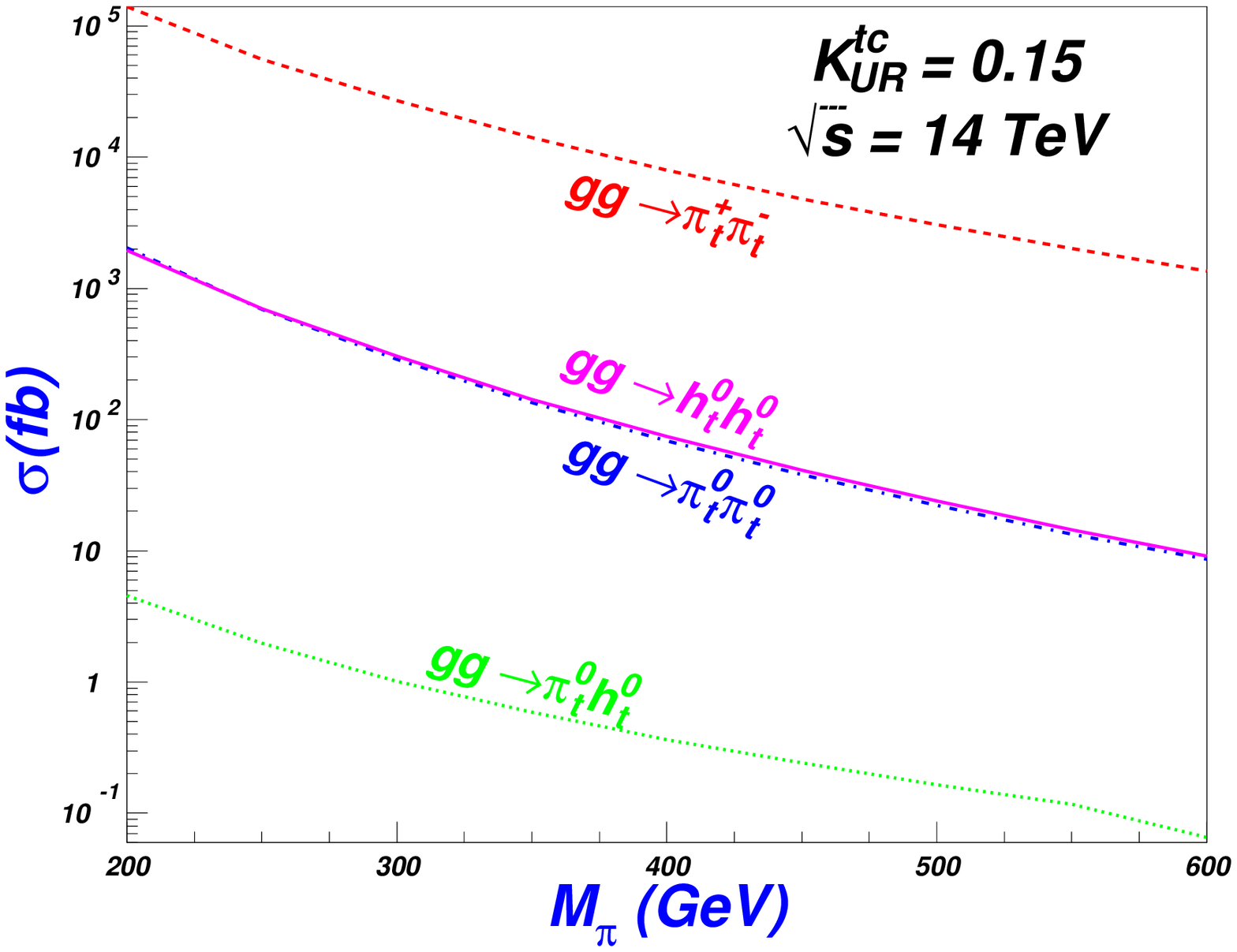,width=8.25cm} }
 \hspace*{0.2cm}
 \parbox{8.05cm}{\epsfig{figure=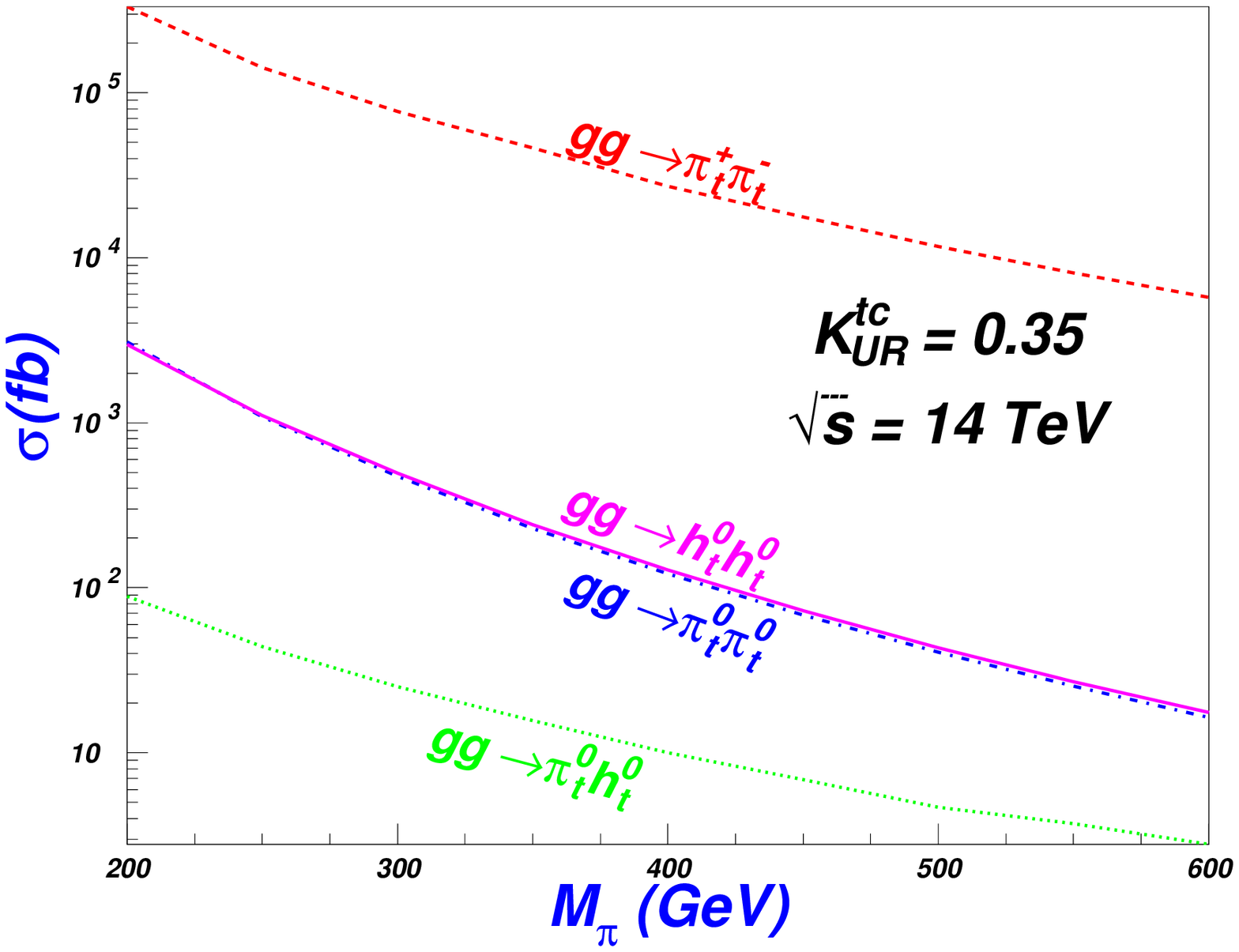,width=8.25cm} }
 \caption{Same as Fig.\ref{ggpp1}, but for $\sqrt{s}=14$ TeV.
\label{ggpp2}  }
\end{center}
\end{figure}

The production cross sections of the $\pi_t^+\pi_t^-$ and $\pi_t^0
h_t^0$ $\pi_t^0 \pi_t^0$, $h_t^0 h_t^0$ of the $gg$ fusion are
plotted in Figs.\ref{ggpp1}, \ref{ggpp2} for $\sqrt{s}=7,~14$ TeV
and $K_{UR}^{tc}= 0.15, ~0.35$, as functions of the top-pion mass
$m_\pi$, assuming the top-higgs mass, $m_h =m_\pi$. From which, we
can see the cross section of this process is quite large, about $1$
pb in most of the parameter space and, as was expected, the
production rate decreases with the increasing top-pion mass since
the phase space are depressed by the top-pion mass.

In Figs.\ref{ggpp1}, \ref{ggpp2} we can also see the $K_{UR}^{tc}$
dependence of the process $gg\to \pi_t^+\pi_t^-$ is very weak since,
in Fig.1 (a), the dominant contribution is the $\pi_t^+ t\bar b $
coupling, which is decided by factor $Y\sim 3$, irrelevant of the
parameter $K_{UR}^{tc}$. Of course, the $\pi_t^+\bar b c $ may also
contribute by $b$, $c$ quarks entering the loop, the production
rate, however, are brought down by the $(K_{UR}^{tc})^4$ since the
vertex $\pi_t^+\bar b c $ appears twice in the loop diagrams. Even
we take the optimum value of $K_{UR}^{tc} \sim 0.4 $, the cross
section will be $1/40$ depressed,  so the cross section contributed
by $bbbc$ and $cccb$ loop is very small. On this backgrounds, the
interference terms between the loops contributed by $t,b$ and $c,b$
are also very small, less than $100$ fb.

Similarly, the gluon gluon fusion processes of the $\pi^0_t\pi^0_t$
and $h^0_th^0_t$ neutral scalar production, may be possess totally
same behaviors since the couplings $\pi_t^0 t\bar t $ or $h_t^0
t\bar t $ are also independent of parameter $K_{UR}^{tc}$ and the
flavor changing couplings $\pi_t^0 t\bar c $ or $h_t^0 t\bar c $
contribute small since this type of couplings appears  twice in the
box diagrams, which can be seen clearly by comparing the
Fig.\ref{ggpp1} and Fig.\ref{ggpp2} by different values of the
parameter $K_{UR}^{tc}$.

But for the $gg\to \pi_t^0h^0_t$ process, the situation will be
different. Since the interaction between the CP-even and the CP-odd
states may cancel out each other, the contributions from the all top
quarks in the box loop may be much smaller than that of the $tttc$
and $ccct$ loop, so the terms with parameter $K_{UR}^{tc}$ may play
a great role. This is also verified by Fig.\ref{ggpp1} and
Fig.\ref{ggpp2}, from which we can see that the the rate of the
$gg\to \pi_t^0h^0_t$ is about two orders smaller than those of the
$gg\to \pi_t^0\pi^0_t$ and $h_t^0h^0_t$, since in the latter the
$tttt$ loop contributes large, and that the cross section of the
$gg\to \pi_t^0h^0_t$ is very sensitive to the $K_{UR}^{tc}$. For
$\sqrt{s} = 14$ TeV, the cross section of the process $gg\to
\pi_t^0h^0_t$ arrives at $88$ fb when $K_{UR}^{tc}=0.35$, but only
$4.6$ fb when $K_{UR}^{tc}=0.15$.


Summarily, For the the parameters $K_{UR}^{tc}$ dependence,  we can
see from the fig.\ref{ggpp1,ggpp2} that the parameter it affects the
rates of the production, and the cross section will increase with
increasing $K_{UR}^{tc}$,  but the effect is not too large. When the
$K_{UR}^{tc}$ increases form $0.15$ to $0.35$, the cross sections
are in the same order. The rates of the $\pi_t^+\pi_t^-$ production
for $\sqrt s = 14 $ TeV, for example, are $210$ pb and $340$ pb, for
$K_{UR}^{tc}=0.15$ and $K_{UR}^{tc}=0.35$, respectively. That is
understandable, since for the charged $\pi_t^+ \pi_t^-$  production,
the $tttb$ and $bbbt$ contribution are primary, which is independent
of the parameter $K_{UR}^{tc}$, while the $cccb$ and $bbbc$ loop are
less important, which is related to the $K_{UR}^{tc}$, directly
proportionally. The same cases occur for the $\pi_t^0\pi_t^0$ and
$h_t^0h_t^0$ neutral production, $tttt$ contribution is larger than
the $tttc$ and $ccct$ ones. But for the $gg\to \pi_t^0h_t^0$, since
the cancellation between the CP-even and CP-odd scalar happen
largely in the $tttt$ loop, the main contributions are from the
$tttc$ and $ccct$ loops, so is is closely connected to the parameter
$K_{UR}^{tc}$.

This discussion are also suitable for the processes $\gamma\gamma
\to SS'$, which we will talk over in Sec. IV. C and the same
conclusion will be talked about very simply.

\subsubsection{$qq\to \pi_t^+ \pi_t^- $ and $qq\to \pi_t^0 h_t^0 $,
 $\pi_t^0 \pi_t^0 $, $h_t^0 h_t^0 $( the last two are for $c\bar c$
 collision) }

\begin{figure}[hbt]
\begin{center}
 \epsfig{file=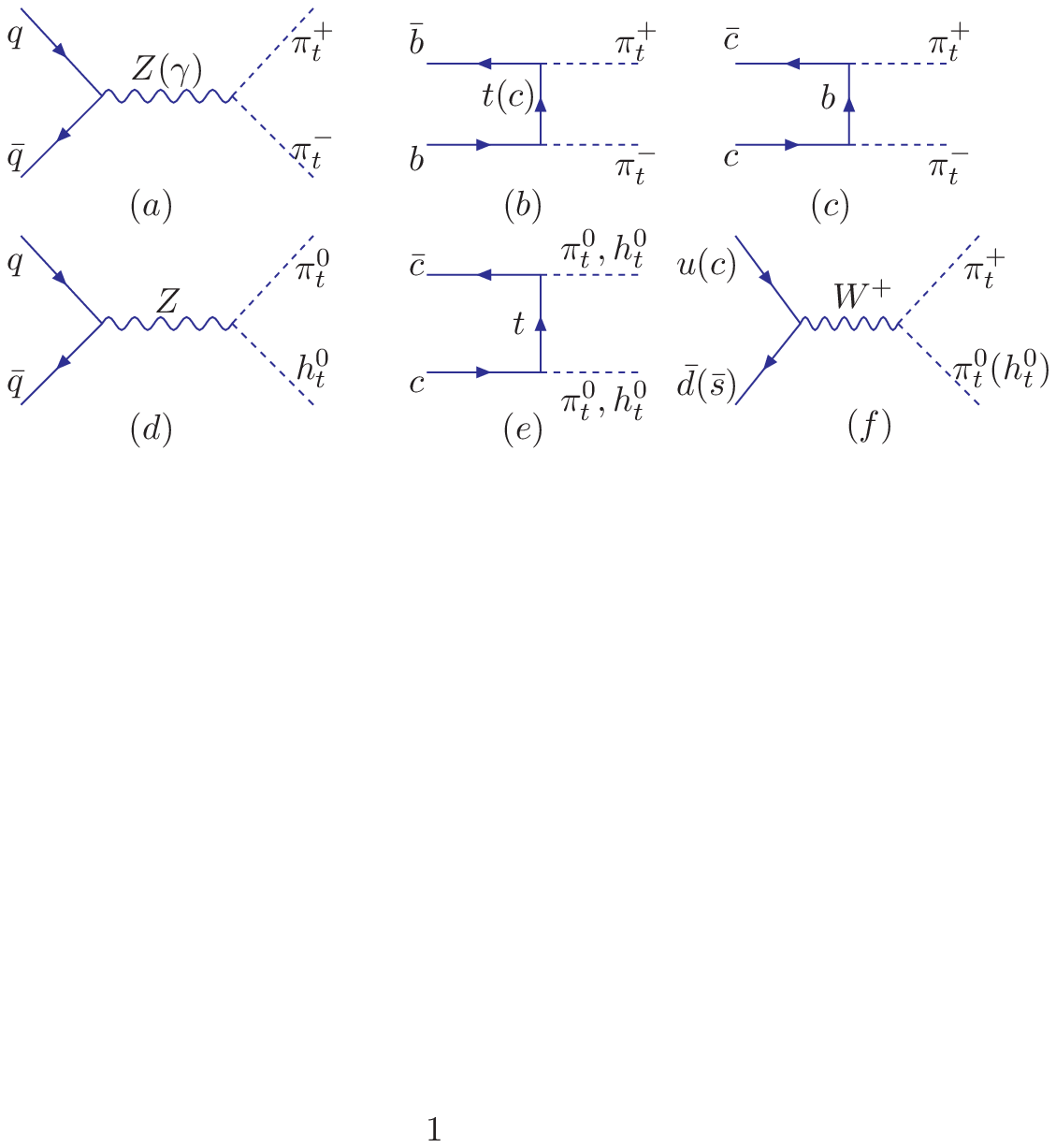,width=12cm}
\caption{Feynman diagrams for the PGB pair production at the LHC via
quark annihilation parton level processes in the TC2 model and
$q=u,d,s,c,b$ quarks.} \label{qqpp-fey}
\end{center}
\end{figure}
\def\figsubcap#1{\par\noindent\centering\footnotesize(#1)}
\begin{figure}[bht]%
\begin{center}
\hspace{-0.8cm}
 \parbox{8.05cm}{\epsfig{figure=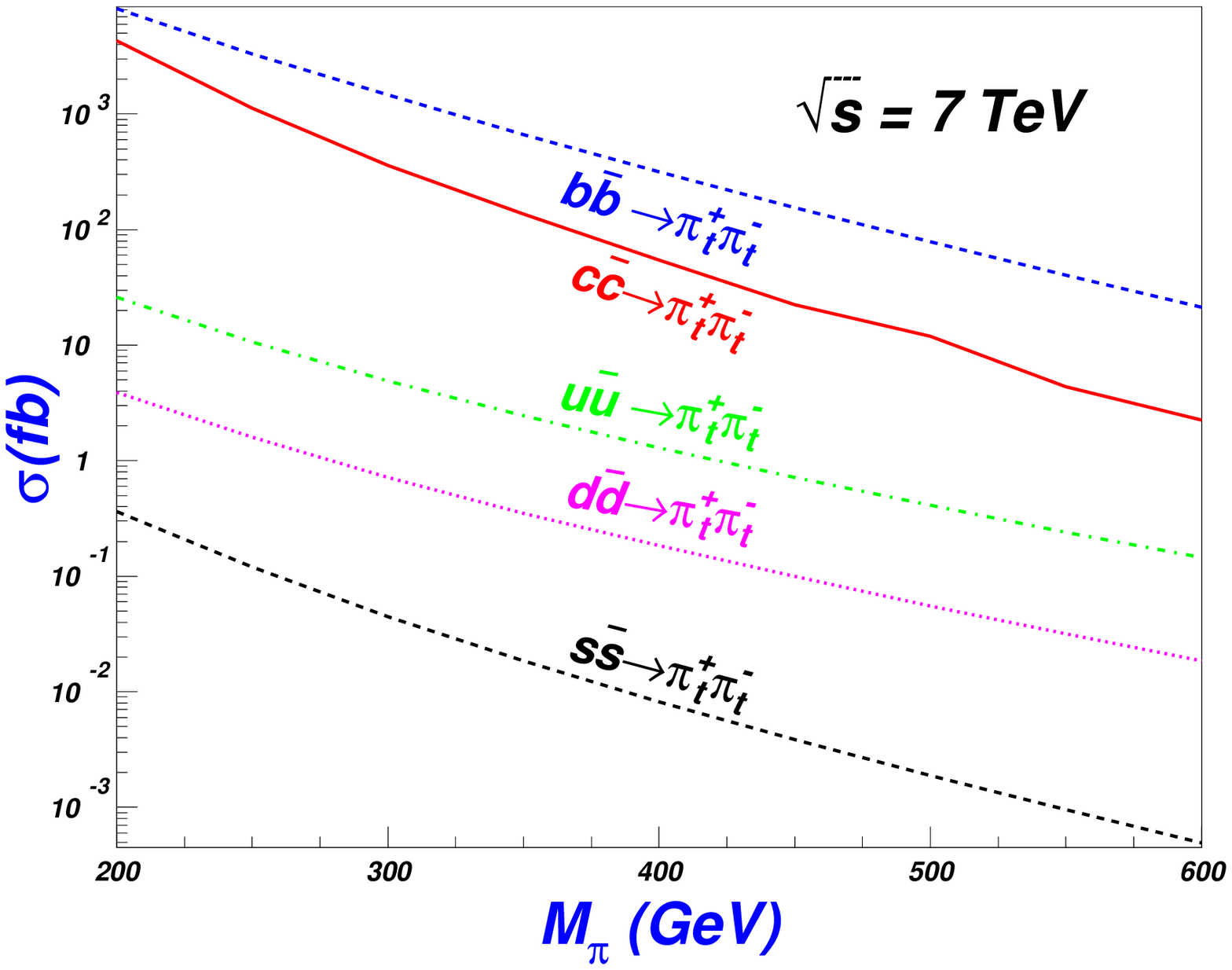,width=8.25cm} }
 \hspace*{0.2cm}
 \parbox{8.05cm}{\epsfig{figure=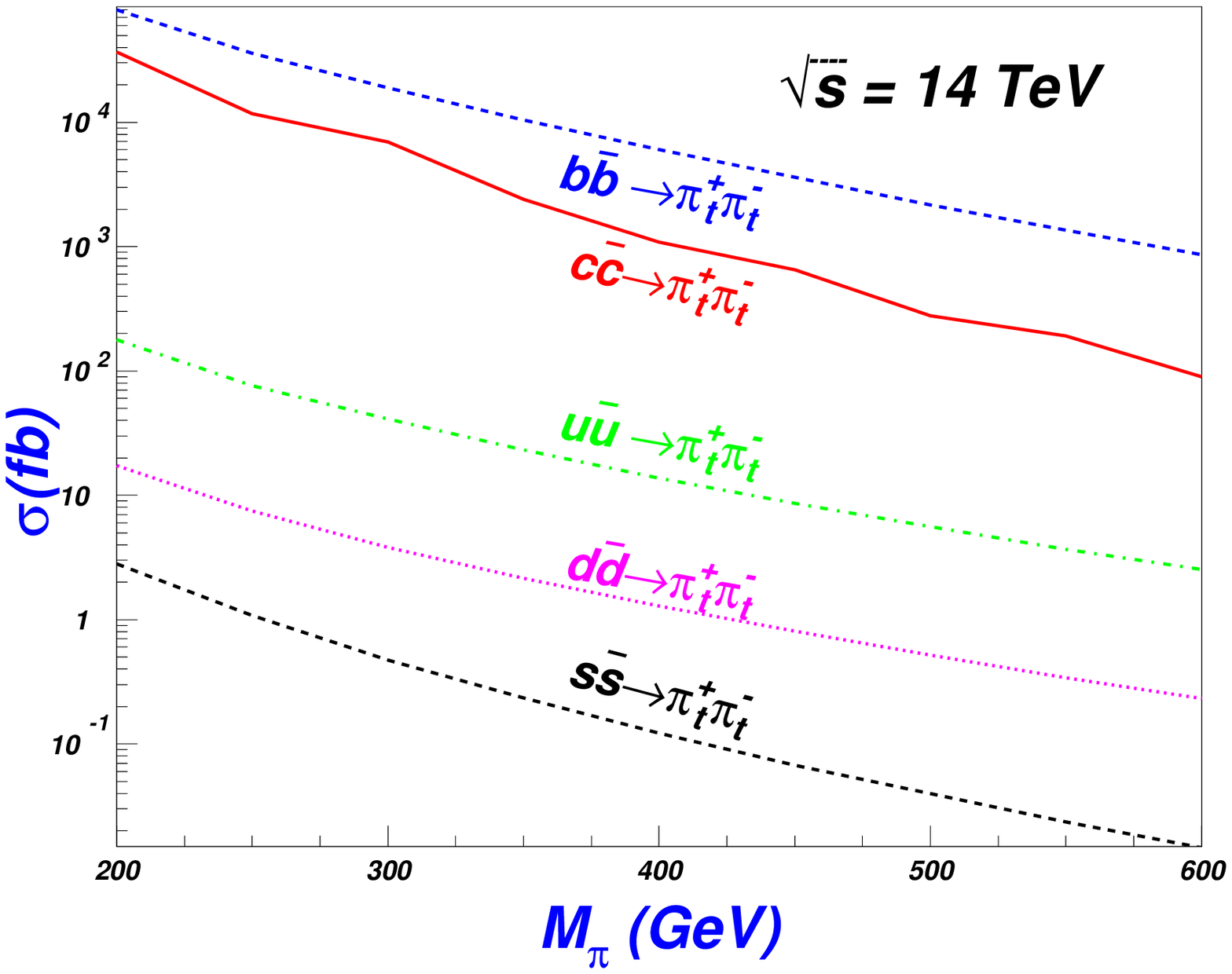,width=8.25cm} }
 \caption{ The cross section $\sigma$ of the processes
 $q\bar q\to \pi_t^+\pi_t^-$ as a function of the top-pion mass $m_{\pi_t}$ with
$\sqrt{s}=7$ TeV  and $\sqrt{s}=14$ TeV, $q=u,d,s,c,b$.
\label{qqpp1} }
\end{center}
\end{figure}
Here, the t-channel neutral production $\pi_t^0 h_t^0 $,
 $\pi_t^0 \pi_t^0 $, $h_t^0 h_t^0$ should have different
cross sections with different couplings, but $\pi_t^0 \pi_t^0 $,
$h_t^0 h_t^0$ production are only contained by the t-channel
processes, so we firstly take $\pi_t^0 h_t^0 $ as an example to
compare with others, and  the t-channel, i.e., $c\bar c \to \pi_t^0
h_t^0$, $\pi_t^0 \pi_t^0 $, $h_t^0 h_t^0$ will appear in the final
of this section.

 The s-channel processes such as (a)(d)(f) in
Fig.~\ref{qqpp-fey}, though the parton distribution functions could
be larger for the $u\bar u$ and $d\bar d$ initial state, may be
relatively small in view of the center-of-mass depression effects.
At the same time, the t-channel coupling strengths are larger than
those of the s-channel. In Fig.~\ref{qqpp-fey}(b), For instance, the
strengthen of $\pi_t^+t\bar b \sim m_t/F_t \sim 3$ is much larger
than that of $Z\pi_t^+\pi_t^- $, $Z\pi_t^0 h_t^0 $ and
$W\pi_t^+\pi_t^0 $ in the s-channel processes, so no wonder the
cross sections of the parton level processes like $u\bar u(d\bar d,
s \bar s) \to Z \to \pi_t^+\pi_t^-(\pi_t^0 h_t^0)$ may be smaller
than those of the others even with larger parton distribution
functions. These can be seen clearly in
Figs.\ref{qqpp1},\ref{qqpp3}. 

 From Fig.\ref{qqpp1},
we can also see that the largest channel of the processes $qq\to
\pi_t^+ \pi_t^- $ is the $b\bar b \to \pi_t^+\pi_t^-$ and $c\bar
c\to \pi_t^+\pi_t^-$, which is easy to understand since, in
Fig.\ref{qqpp1}, the t-channel processes (b) and (c) are free of the
center-of-mass depression and larger than others. The former, i.e.,
the process $b\bar b \to \pi_t^+\pi_t^-$, however, surpasses the
process $c\bar c  \to \pi_t^+\pi_t^-$, since the vertex
$\pi_t^+t\bar b $, different from $\pi_t^+ c\bar b $, is not
associated with the $K_{UR}^{tc}$ and not reduced by it.

But for the neutral scalar production via the $q \bar q$ collision,
there is only one t-channel contribution in Fig.\ref{qqpp-fey} (e)
since flavor changing neutral couplings induced by the neutral
scalars $\pi_t^0$ and $ h_t^0$ are small $\sim m_q$($m_q$ is the
quark mass) \cite{tc2-rev}, except the $\pi_t^0 (h_t^0)t\bar c$
coupling $\sim m_t$ with the large top quark mass, which appears in
the t-channel of the $c\bar c \to \pi_t^0 h_t^0$.

For the neutral scalar production induced by the $q \bar q$
collision, the t-channel contribution is the largest when the
production rates are not depressed too much by the factor
$K_{UR}^{tc}$, about $100$ fb and the other processes are smaller
and have different cross sections.  What makes the difference among
them is only, if we neglect the masses of the quarks $u,d,c,s, b$,
the parton distribution function in the proton, so it is naturally
to see that $\sigma(u\bar u) >\sigma(d\bar d) >\sigma(s\bar s)
>\sigma(b\bar b) $.

\begin{figure}[bht]%
\begin{center}
\hspace{-0.8cm}
 \parbox{8.05cm}{\epsfig{figure=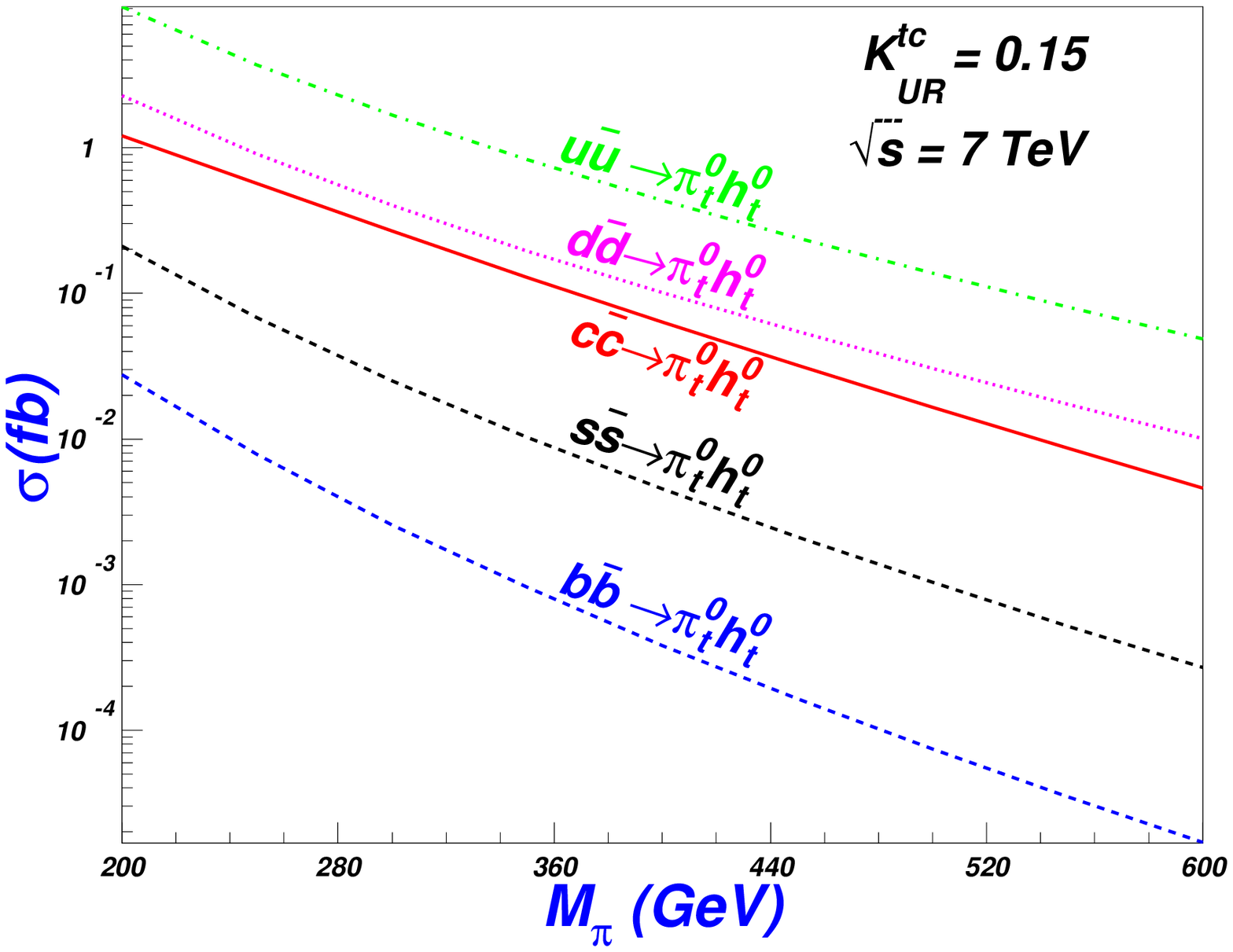,width=8.25cm} }
 \hspace*{0.2cm}
 \parbox{8.05cm}{\epsfig{figure=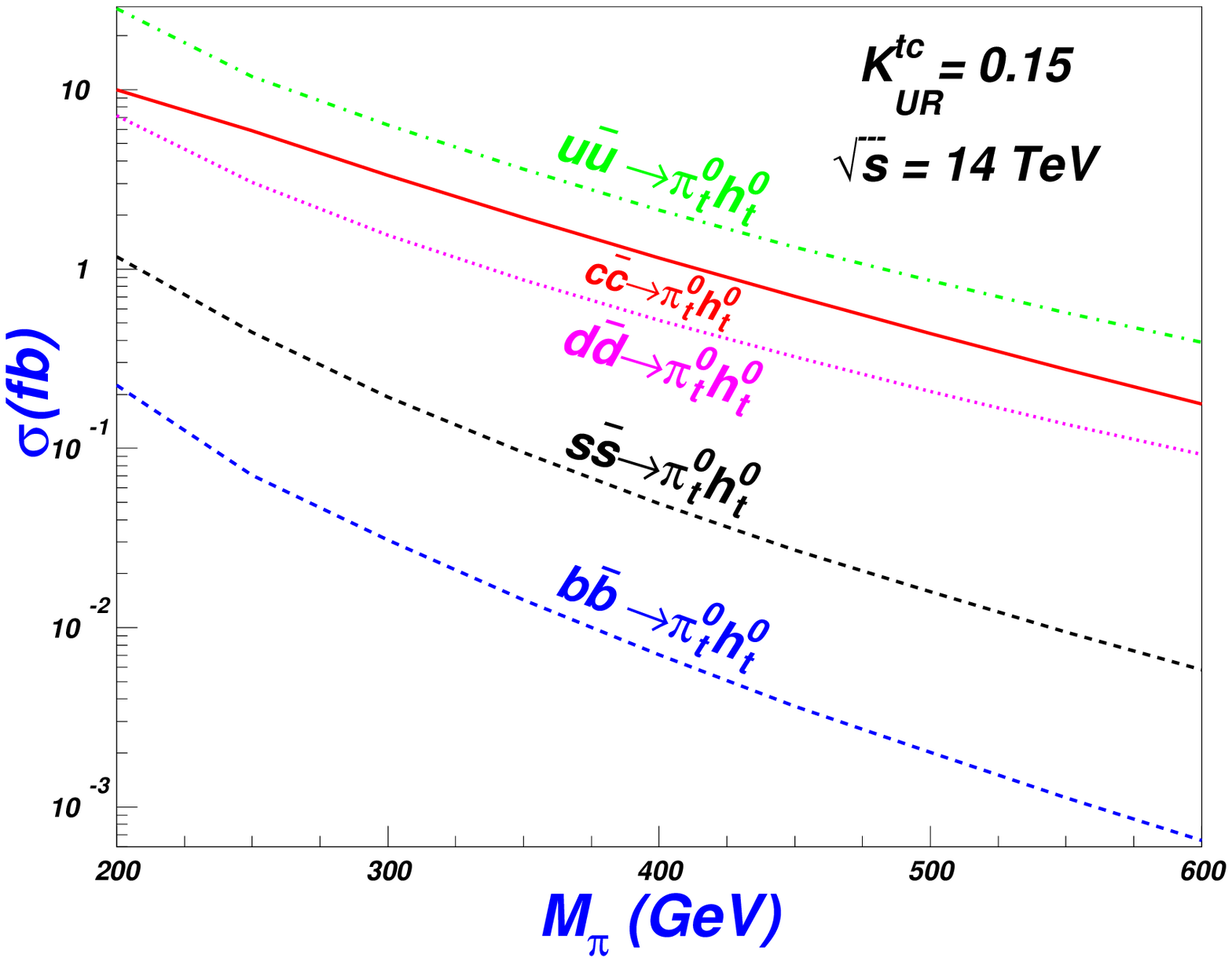,width=8.25cm} }r
 \caption{ The cross section $\sigma$ of the processes
 $q\bar q\to \pi_t^0
h_t^0 $ as a function of the top-pion mass $m_{\pi_t}$ with
 $K_{UR}^{tc}=0.15$ and $K_{UR}^{tc}=0.35$ and for $\sqrt{s}=7$ TeV, $q=u,d,s,c,b$. \label{qqpp2} }
\end{center}
\end{figure}

\begin{figure}[bht]%
\begin{center}
\hspace{-0.8cm}
 \parbox{8.05cm}{\epsfig{figure=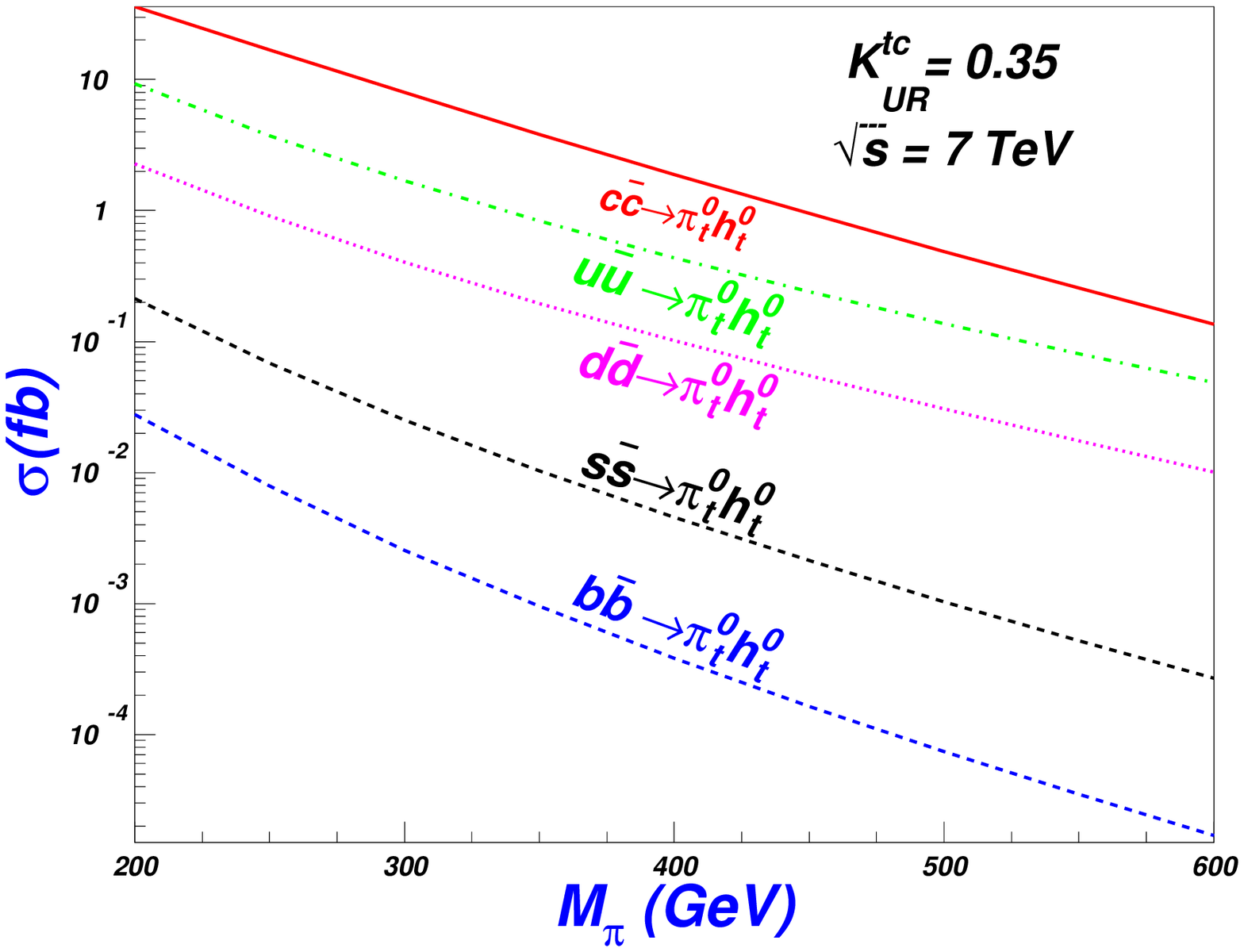,width=8.25cm} }
 \hspace*{0.2cm}
 \parbox{8.05cm}{\epsfig{figure=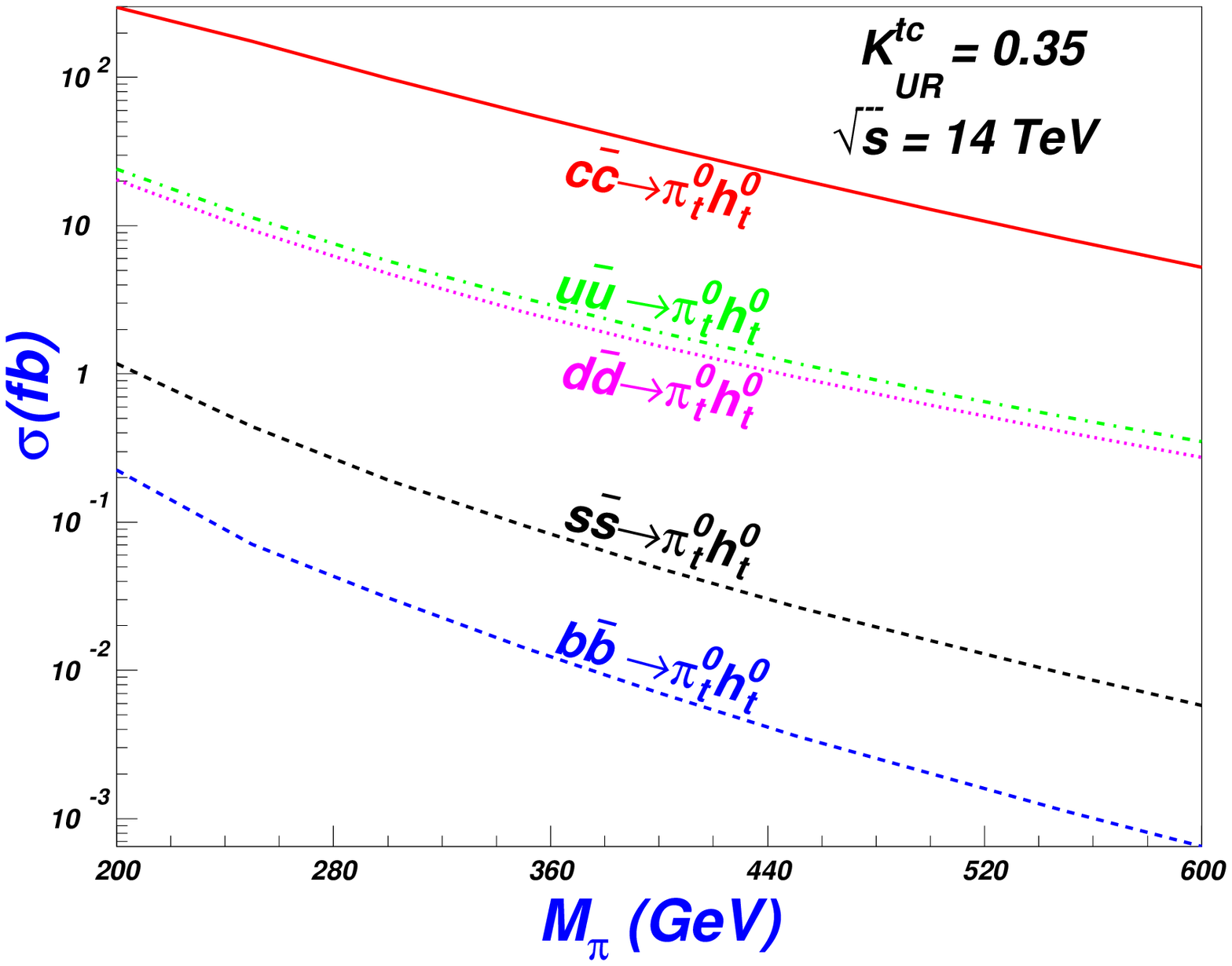,width=8.25cm} }
 \caption{ Same as Fig.\ref{qqpp1}, but for $\sqrt{s}=14$ TeV.
\label{qqpp3} }
\end{center}
\end{figure}

Figs.\ref{qqpp2}, \ref{qqpp3} also shows $m_\pi$ dependence of the
cross section for $K_{UR}^{tc} = 0.15$ and $0.35$, respectively.
Comparing Fig.\ref{qqpp1} and Figs.\ref{qqpp2}, \ref{qqpp3},  we can
see that, in the latter, the cross sections becoming smaller,
especially the t-channel processes $c\bar c \to \pi_t^+ \pi_t^- $
and $c\bar c \to \pi_t^0 h_t^0$, about $1/30$ of the former. That is
easy to understand since the in the amplitudes the $\pi_t^+ b \bar c
$ and $\pi_t^0 t \bar c $ or $h_t^0 t \bar c$ vertex, $\sim
K_{UR}^{tc}$ appears twice, so the cross sections decrease
$(0.15/0.35)^4 \sim 1/30$. But for process $b\bar b \to \pi_t^+
\pi_t^-$, when we take $K_{UR}^{tc} = 0.15$, the depression rate is
only $1/2$, that is because in this process, when the intermediate
particle is top quark in Fig.\ref{qqpp-fey}(b), the contribution is
dominant and the coupling $\pi_t^+ t \bar b\sim \frac
{m_t}{F_t}\frac{\sqrt{V_W^2-F_t^2}}{V_W}$ is irrelevant to the
depression parameter $K_{UR}^{tc}$, so the rate is not sensitive to
it too much.

Note that, for the neutral scalar productions, in Fig.\ref{qqpp2},
$\sigma(u\bar u \to \pi_t^0 h_t^0) >\sigma(c\bar c \to \pi_t^0
h_t^0)$, which is opposite to the situation of Fig.\ref{qqpp3}. This
also shows that the $c\bar c \to \pi_t^0 h_t^0$ decreases with the
decreasing $K_{UR}^{tc}$ and more simultaneously, the process $u\bar
u \to \pi_t^0 h_t^0$ is not related to the parameter $K_{UR}^{tc}$.
Actually, all the s-channel processes in Fig.\ref{qqpp-fey}(a),(d),
are immune to the depression parameter $K_{UR}^{tc}$. So the weak
advantage of the $\sigma(c\bar c \to \pi_t^0 h_t^0)$ over
$\sigma(u\bar u \to \pi_t^0 h_t^0) $ will fade away with the
decreasing $K_{UR}^{tc}$.
\def\figsubcap#1{\par\noindent\centering\footnotesize(#1)}
\def\figsubcap#1{\par\noindent\centering\footnotesize(#1)}

We also see from Fig.\ref{qqpp1} and Figs.\ref{qqpp2},\ref{qqpp3}
that the charged scalar pair productions are much larger than the
neutral ones with the same parameters, i.e, the top-pion mass
$m_\pi$ and $K_{UR}^{tc}$. The rate of $c\bar c \to \pi_t^+\pi_t^-$,
for example, is about two orders larger than that of the $c\bar c
\to \pi_t^0 h_t^0$, which is simple to understand since the $\pi_t^+
t \bar b $ is free of the $K_{UR}^{tc}$ depression and for
$\pi_t^0(h_t^0) t \bar c $ that is not the truth.

We here only discussion the neutral pair production $\pi_t^0h_t^0$,
while for the $\pi_t^0\pi_t^0$ and the $h_t^0h_t^0$ production, the
final particles are identical particles, due to identical particle
statistics, the cross section of them would each be equal to
$(1/2)^2$ of the $\pi^0_t h_t^0$ cross section with the same scalar
masses, considering the same coupling strength.
\def\figsubcap#1{\par\noindent\centering\footnotesize(#1)}
\begin{figure}[bht]%
\begin{center}
\hspace{-0.8cm}
 \parbox{8.05cm}{\epsfig{figure=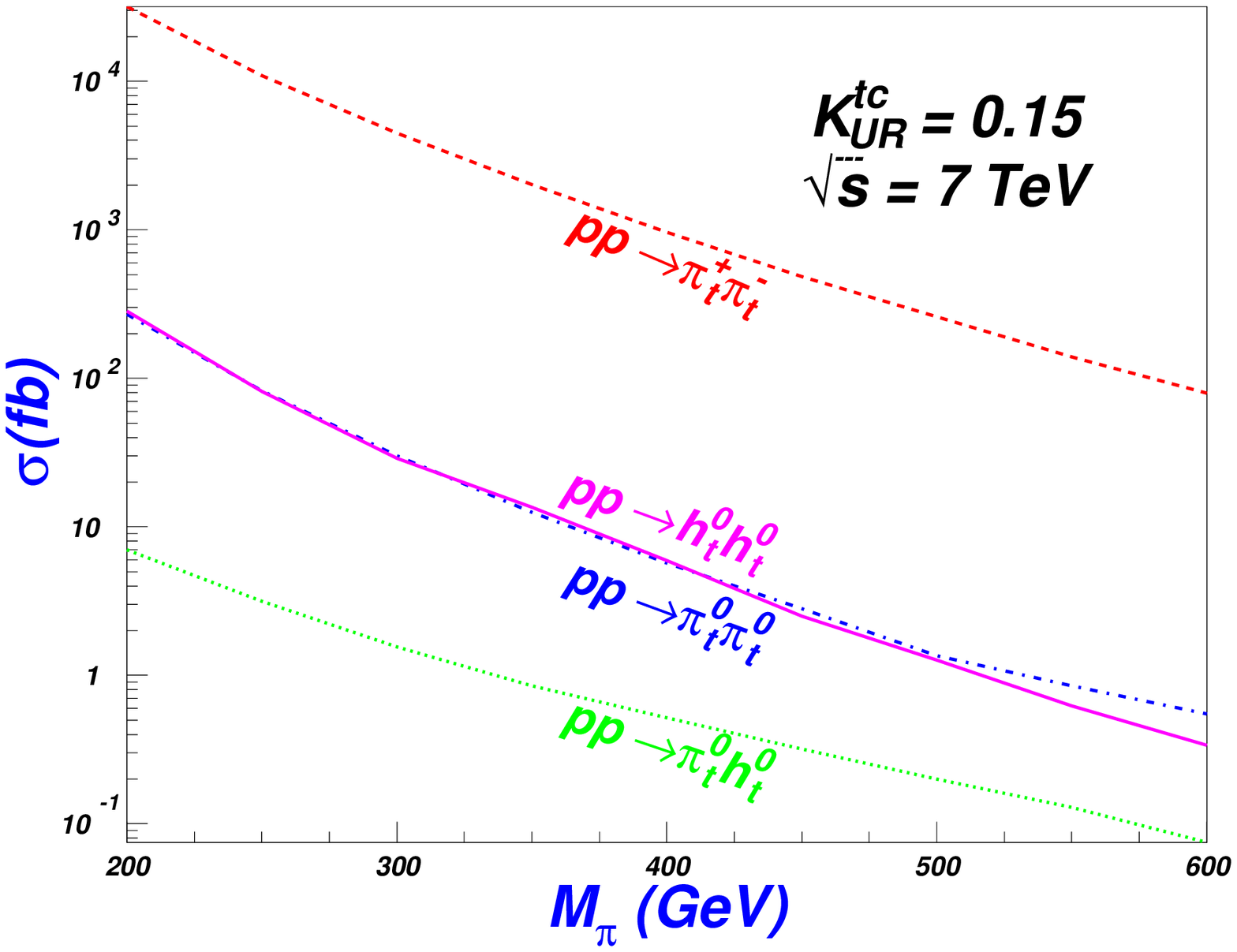,width=8.25cm}
 \figsubcap{a}}
 \hspace*{0.2cm}
 \parbox{8.05cm}{\epsfig{figure=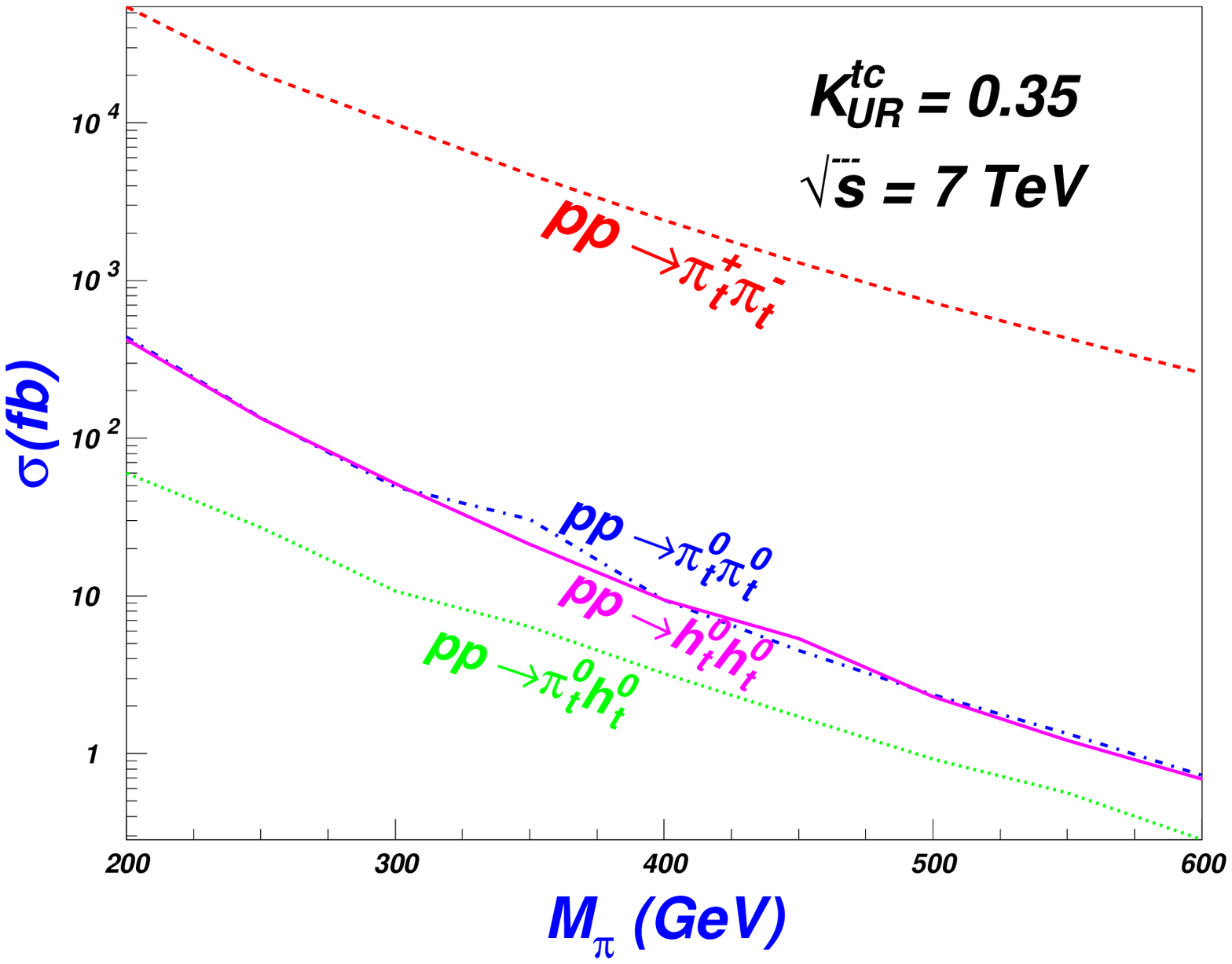,width=8.25cm}
 \figsubcap{b}}
 \caption{ The total cross section $\sigma$ of the processes
 $p\bar p\to SS'$  as a function of the
top-pion mass $m_{\pi_t}$ with
 $\sqrt{s}=7$ TeV and for  $K_{UR}^{tc}=0.15, ~0.35$. \label{total-pp1} }
\end{center}
\end{figure}

\def\figsubcap#1{\par\noindent\centering\footnotesize(#1)}
\begin{figure}[bht]%
\begin{center}
\hspace{-0.8cm}
 \parbox{8.05cm}{\epsfig{figure=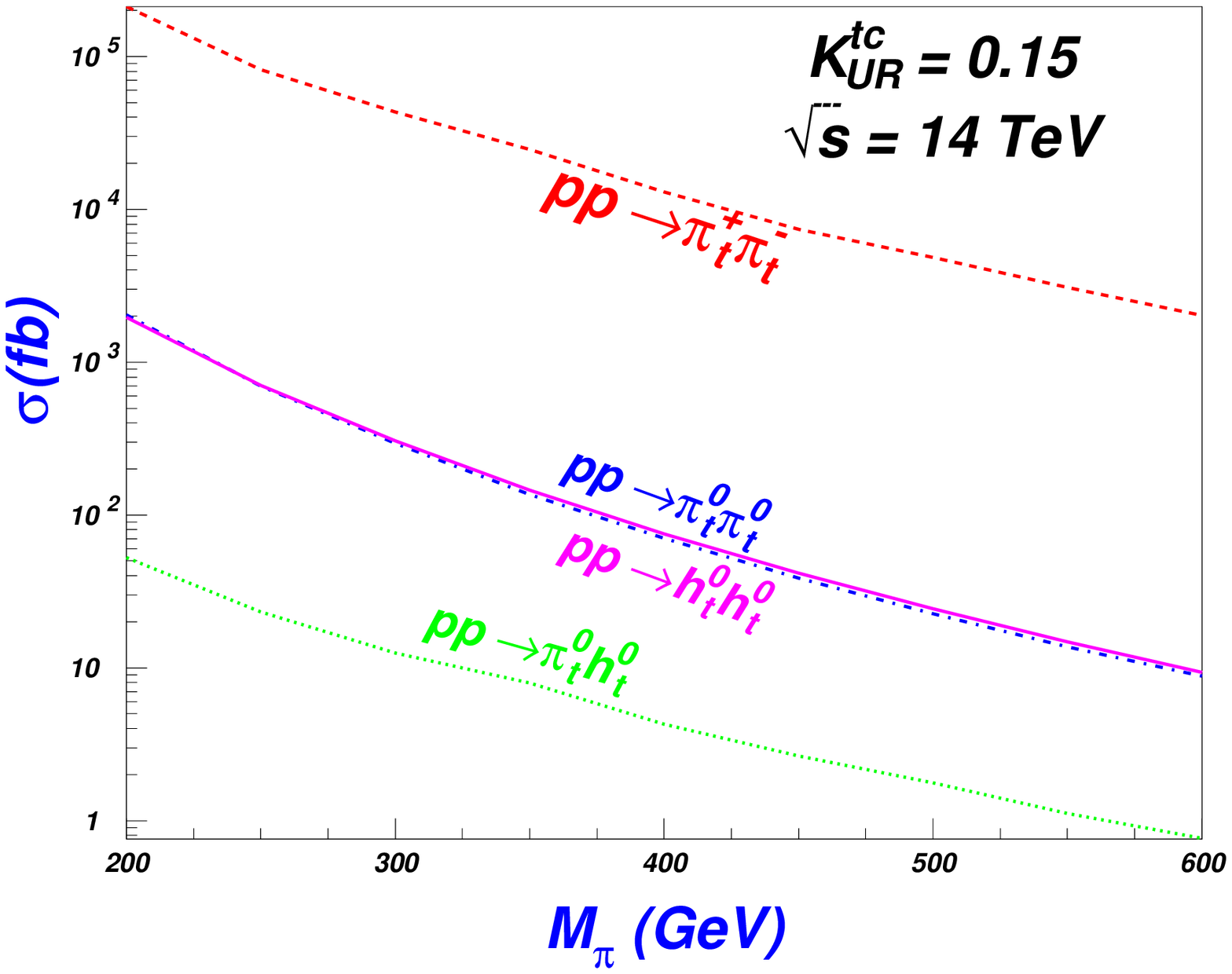,width=8.25cm}
 \figsubcap{a}}
 \hspace*{0.2cm}
 \parbox{8.05cm}{\epsfig{figure=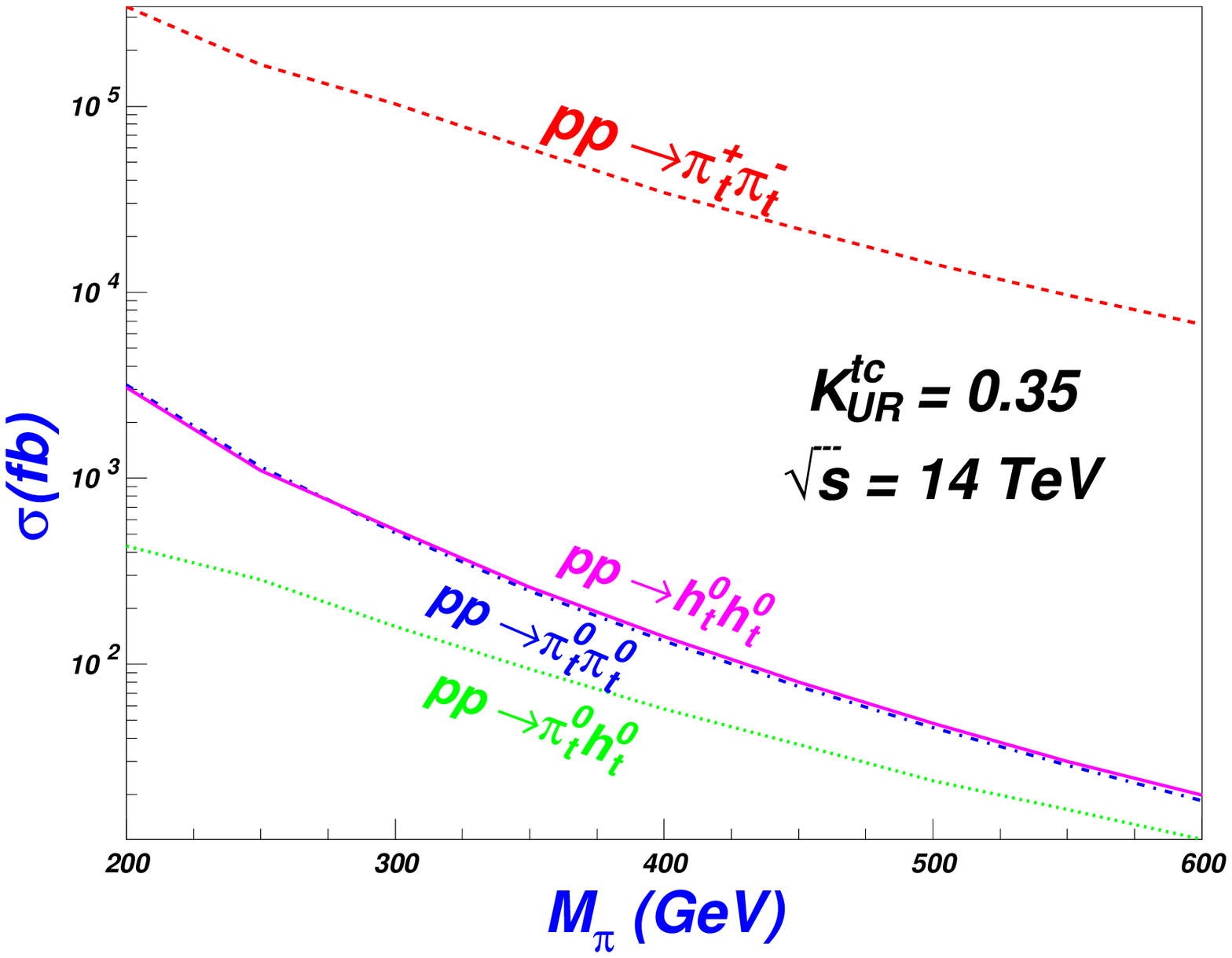,width=8.25cm}
 \figsubcap{b}}
 \caption{ Same as Fig.\ref{total-pp1}, but for $\sqrt{s}=14$ TeV. \label{total-pp2} }
\end{center}
\end{figure}

\subsubsection{The total contribution for the $\pi_t^+ \pi_t^-$ and $\pi_t^0
h_t^0$, $\pi_t^0 \pi_t^0$, $h_t^0 h_t^0$ at the LHC}

Here we sum all the contributions, just shown as Fig.\ref{total-pp1}
and Fig.\ref{total-pp2}. From which we can see the total production
rate of the charged top-pions is related to the top-pion mass and
the center-of-mass and the production probability is larger than
$6709$ fb when the center-of-mass $\sqrt s = 14 $ TeV and larger
than $79$ fb when $\sqrt s = 7 $ TeV for $m_\pi = 600 $ GeV. While
for the neutral production of $pp\to h_t^0h_t^0$ and
$\pi_t^0\pi_t^0$, the cross sections are about 2-3 orders smaller
than the charged one. The cross section of the $h_t^0h_t^0$ final
state, for example, is about $3$ pb for $m_\pi = 200 $ GeV and
$\sqrt s = 14 $ TeV, while for the charged one, the rate can arrive
at $340$ pb.

  We also see that the rates of the processes $pp\to h_t^0h_t^0$ and
$\pi_t^0\pi_t^0$, are $1-2$ orders larger than that of the $pp\to
\pi_t^0h_t^0$, which is because, in these productions, the $gg$
fusion contributes most, while $q\bar q$ collsion does not change
the trend.

We can also see that $K_{UR}^{tc}$ dependence is also almost the
same as that of the $gg$ fusion for every channel, which proves
again that the the $gg$ fusion contributes largely.

\subsubsection{$u\bar d (c\bar s) \to \pi_t^+ \pi_t^0(h_t^0) $ }
 \def\figsubcap#1{\par\noindent\centering\footnotesize(#1)}
 \begin{figure}[bht]%
 \begin{center}
  \parbox{8cm}{\epsfig{figure=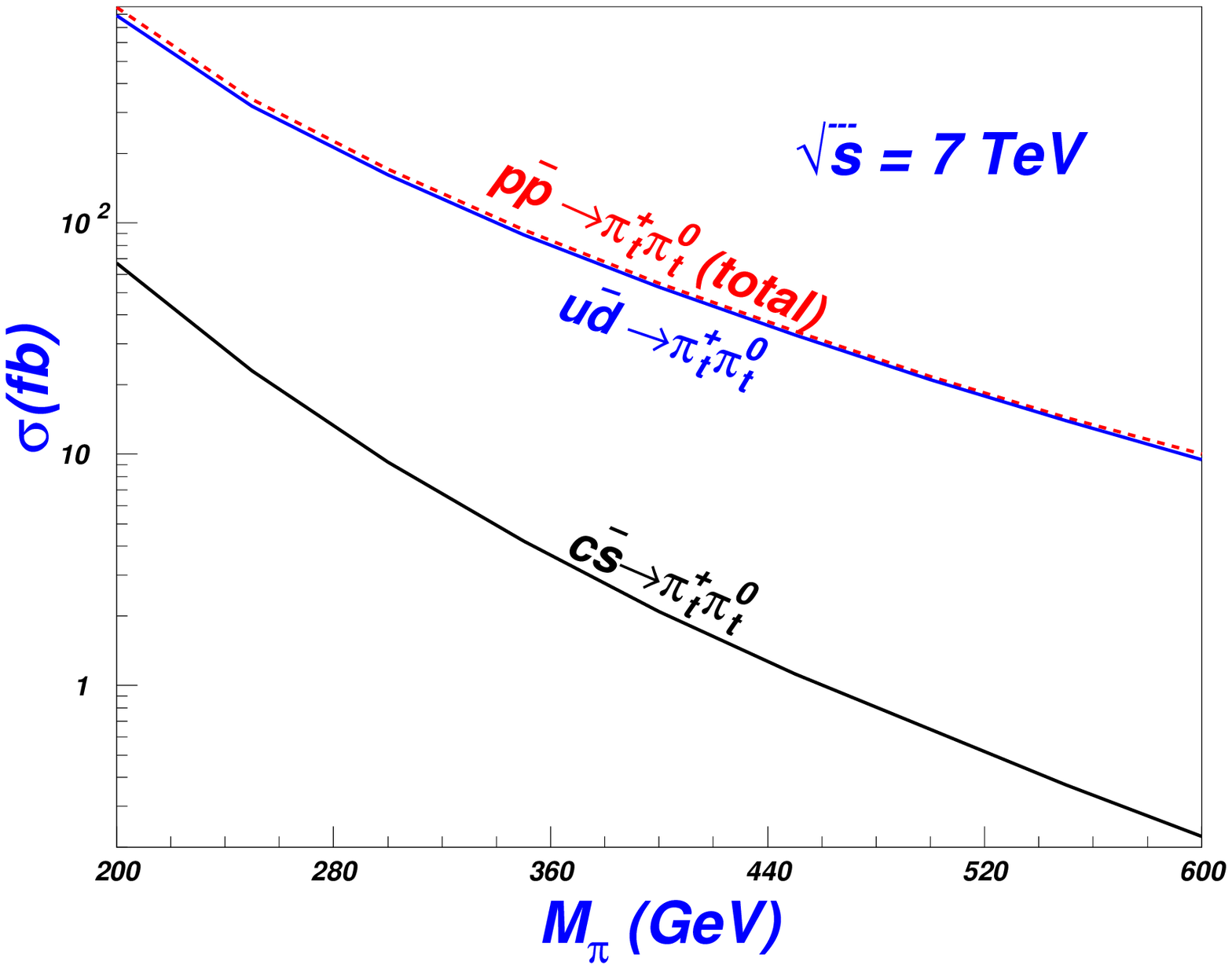,width=8cm}  }  
   \parbox{8cm}{\epsfig{figure=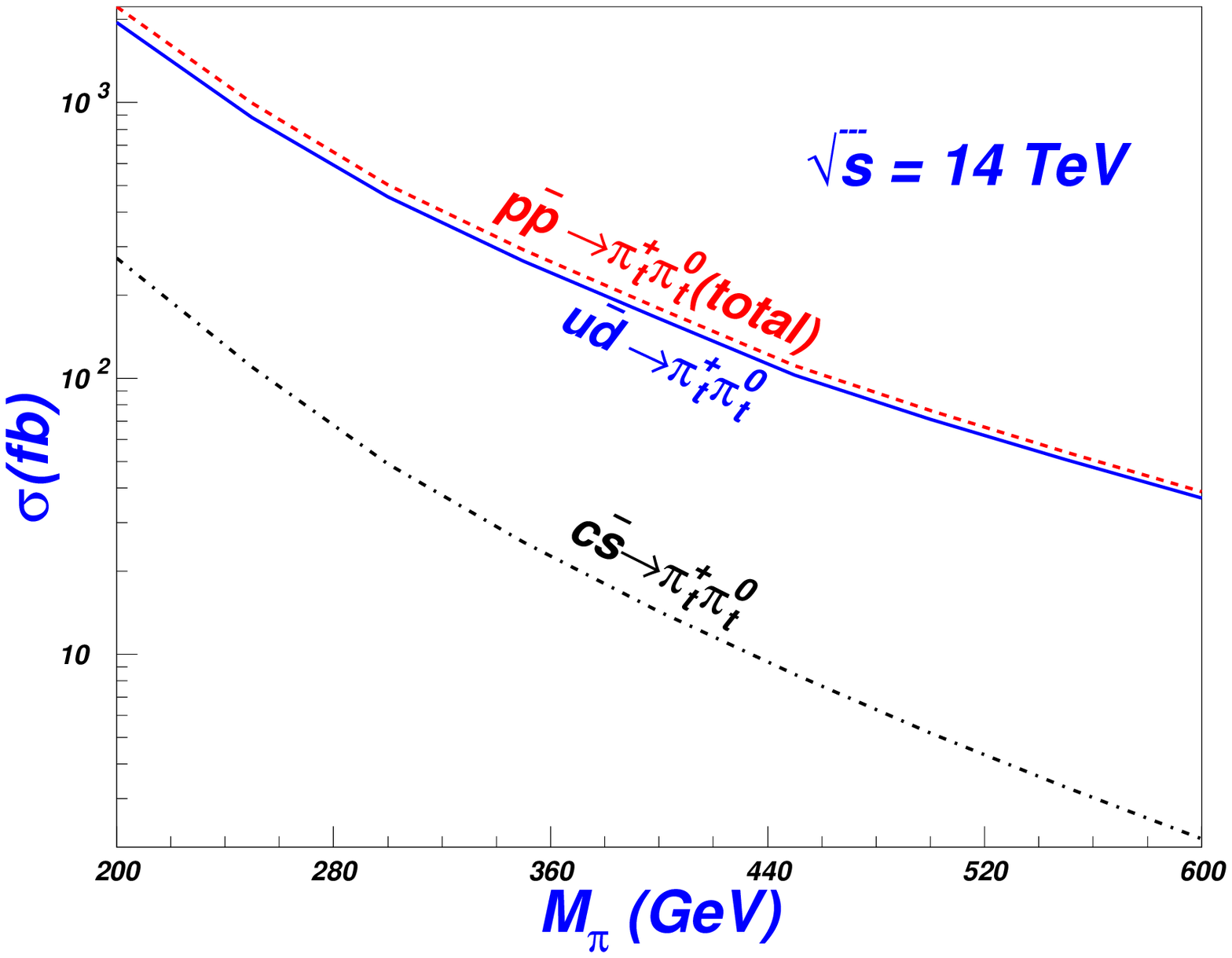,width=8cm} }  
 \caption{ The cross section $\sigma$ of the processes $u\bar d\to \pi_t^+\pi_t^0$
 and $c\bar s\to \pi_t^+\pi_t^0$ as
a function of the top-pion mass $m_{\pi_t}$ with $\sqrt{s}=7$ TeV
and $\sqrt{s}=14$ TeV . \label{udp+h0} }
 \end{center}
 \end{figure}

From Table~\ref{tbl:coups1}, we find the $W^{-\mu} h_t^0\pi_t^+$ and
$W^{-\mu} \pi_t^0\pi_t^+$ couplings, which makes the one charged
scalar and one neutral scalar, i.e,  $\pi_t^+ \pi_t^0(h_t^0)$,
associated production possible at the LHC. The difference of the
coupling strength of the $W^{-\mu} h_t^0\pi_t^+$ and $W^{-\mu}
\pi_t^0\pi_t^+$ is only that in  the $\pi_t^+ \pi_t^0(h_t^0)$
coupling, there is an extra $v_T/v$ factor, with $v_T = 241$ GeV and
$v= 246$ GeV, the electroweak scale. So the coupling strength is
almost the same, and the production rates of the $ h_t^0\pi_t^+$ and
$\pi_t^0\pi_t^+$ are taken as the same.
 From Fig.\ref{udp+h0}, we
can see the cross section can arrive to thousands of fb in most of
the parameter spaces. Considering the special final states, this
production may be interesting. We sum all the contributions and
compare them together with that of every channel and find, from
Fig.\ref{udp+h0}, that the $u\bar d$ initial state contributes vast
majority of the total contribution (the sum of that of the $u\bar d$
and the $c\bar s$) so that the two curves of the total and the
$u\bar d \to \pi_t^+ \pi_t^0(h_t^0)$ almost coincide with each
other, which is easy to understand since the parton distribution
function for the first generation is much larger than the others.

\subsubsection{Backgrounds Analysis at the LHC}

For final state $\pi_t^+\pi_t^-$ at the LHC, the charged top-pions
$\pi_t^+$ decays to $t\bar b$ and $c\bar b$ with the branching ratio
about $70\%$ and $30\%$\cite{toppion-decay}, respectively. We assume
the top-pions decaying to $t\bar b$, and top quark to $b$ quark,
charged lepton and the missing energy, i.e. the $4b+2l+\E_slash$
signal\footnote{Actually, usually only 2 bottom quarks are tagged,
so the signal is $2b+2l+2j+\E_slash$.} with $\E_slash$, the missing
energy, so the mainly SM backgrounds are $pp \to WWZjj$(with $Z$ to
$b\bar b$), $WWZZ$(with one $Z$ to $b\bar b$, the other to $jj$),
$WWhh$, $t\bar t W$ (with $W$ to two jets) and $WWb\bar bjj$, where
$h$ decays to $b\bar b$ and the $W\to l\E_slash$. Of course, the
signal cross sections would be reduced by the branching ratios, $
70\%\times 70\%\times 1/6\times 1/6$.

The background cross-sections of the first three processes, i.e,
$WWZjj$, $WWZZ$ and  $WWhh$ are quite small since there are more
than $3$ QED vertexes which depress the strength. Considering the
branching ratio of $W$ and $Z$, the cross sections are at the level
of several tens of fb, so they are negligible in the background
discussion.  For $pp\to t\bar t W$, the production rate, about $500$
fb, similarly, the branching ratio of $W$ decaying to hadrons,
$1/3$, $t\to l\E_slash b$, $1/6$, then signal is about $4.6$ fb,
which is small contrast to the signal. As for the process $pp\to
WWbbjj$, quite large, about $437$ pb, multiplying by the $W$
branching ratios, $12$ pb. To depress it, we apply, first, we can
ask the transverse momentum cut $p_T^j> 20$ Gev, since in the
signal, the transverse momentum of the jets, which are from the
top-pion, are large, while the transverse momentum of the jets  in
the production $pp\to WWbbjj$, are much smaller. So the background
will be cut down largely, without losing much signatures at the same
time. Secondly, the top-pion mass top quark mass reconstruction will
be powerful to depress the background since in the signal the $Wb $
comes form the top quark while in the background, it is not the true
case. For these two means, we believe that the signal will not be
reduced too much, such as  $80\%$ preserving, while the background
may be depressed very much. we, based on the discussion above, here
draw the conclusion that the signal cross sections arriving at
$1000$ fb may be observable at the LHC. Nevertheless, the discussion
here is so crudely, and the precision are far beyond control. We
will, in the next work, debate the observability at length.

Another final states at the LHC, $\pi_t^0 h_t^0$, $\pi_t^0 \pi_t^0$
and $h_t^0 h_t^0$, have same signature since the neutral bosons
$\pi_t^0$ and the $h_t^0$ decay to the same states. If we assume the
two final scalars decaying to $t\bar c $,  the semileptonic decay of
both top (or anti-top) quarks give rise to a signal of like-sign
dilepton plus two b-jets, i.e., $\ell^\pm \ell^\pm + 2$ b-jets
($\ell=e,\mu$), so the signal is like-sign dilepton plus two b-jets,
i.e., $\ell^\pm \ell^\pm + 2$ b-jets + 2 jets($\ell=e,\mu$). Since
we assume only two bottom quarks are tagged, the signal is the same
as the charged top-pion pair production. Therefore the neutral and
charged scalar pair production have same SM backgrounds. So is the
charged and the neutral associated production $\pi_t^+ h_t^0$ and
$\pi_t^+ \pi_t^0$.

Especially, the neutral top-pion or top-higgs pair final states can
yield a like-sign dilepton signal, which is very exciting.  To be
specific, the flavor-changing decay of $\pi^0_t$ or $h^0_t$ will
lead $1/2$ to $t \bar c$ and $1/2$ to $\bar t c$, so that the
neutral pair leptonic decays will be $25\%$ $l^+l^+$, $25\%$
$l^-l^-$, and $50\%$ $l^+l^-$.  This is exciting because the
dominant $t \bar t j j$ background has only opposite-sign leptons.

 To draw a very crudely conclusion, for an integrated luminosity $100$ fb$^{-1}$ at
  the LHC, the scalar pair production cross sections of $1000$ fb may be
  the lower limit of the observability.

\subsection{$e^+e^- \to \pi_t^+\pi_t^-$ and
$e^+e^- \to \pi_t^0 h_t^0$ at the ILC}

\begin{figure}[hbt]
\begin{center}
 \epsfig{file=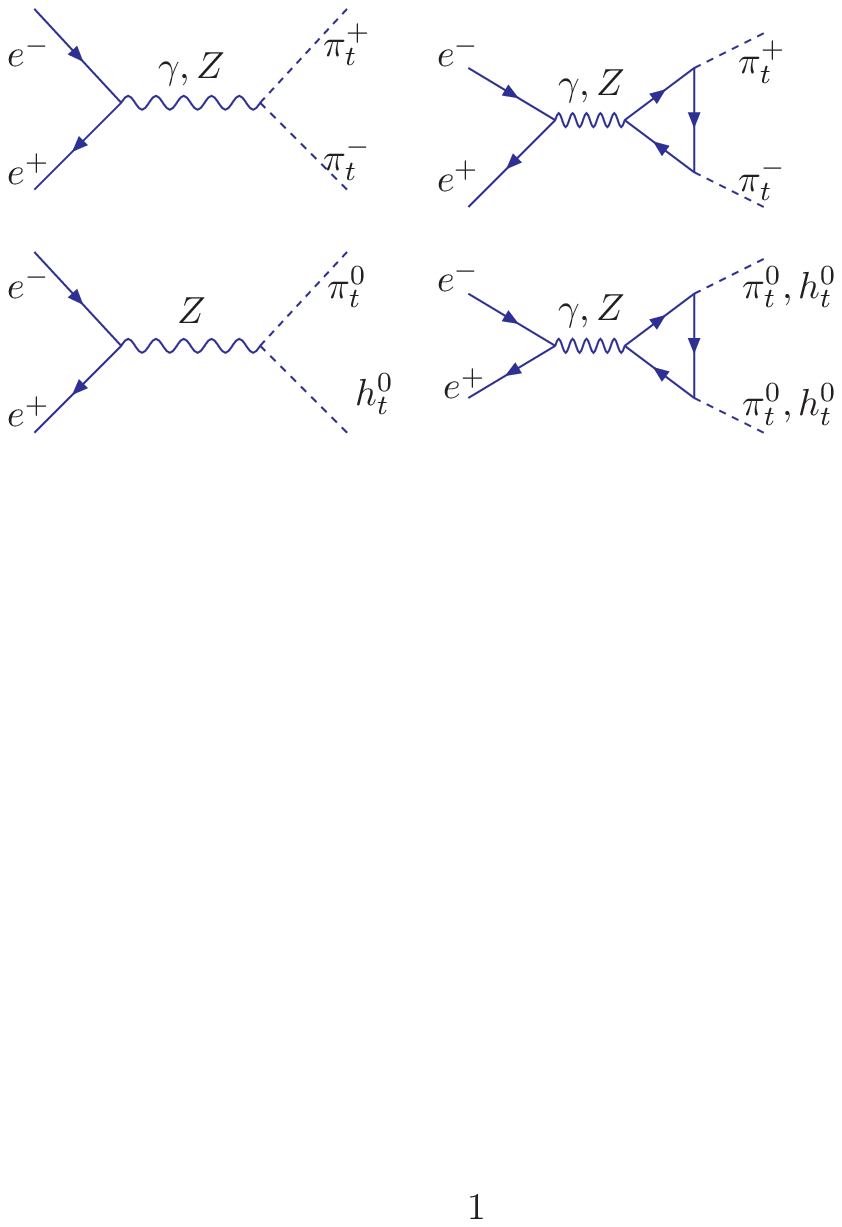,width=12cm}
\caption{Feynman diagrams for the PGB pair production at the ILC via
eletron positron collision processes in the TC2 model.}
\label{eepp-fey}
\end{center}
\end{figure}

At the ILC, PGB pair production carried on by the processes $e^+e^-
\to \pi_t^+\pi_t^-$ and $e^+e^- \to \pi_t^0 h_t^0$, the feynman
diagrams of which are shown in Fig.\ref{eepp-fey}.

The advantage of analyzing such processes at the ILC is obvious that
the hadronic background is very suppressed and the amount of signals
may be practically observable. The calculation of the production at
the $e^+e^-$ collision is relatively simple compared to the case for
hadron colliders because there is no QCD correction and moreover,
there does not exist the complicated infrared divergence which needs
to be properly dealt with.

\def\figsubcap#1{\par\noindent\centering\footnotesize(#1)}
\begin{figure}[bht]%
\begin{center}
\hspace{-0.8cm}
 \parbox{8.05cm}{\epsfig{figure=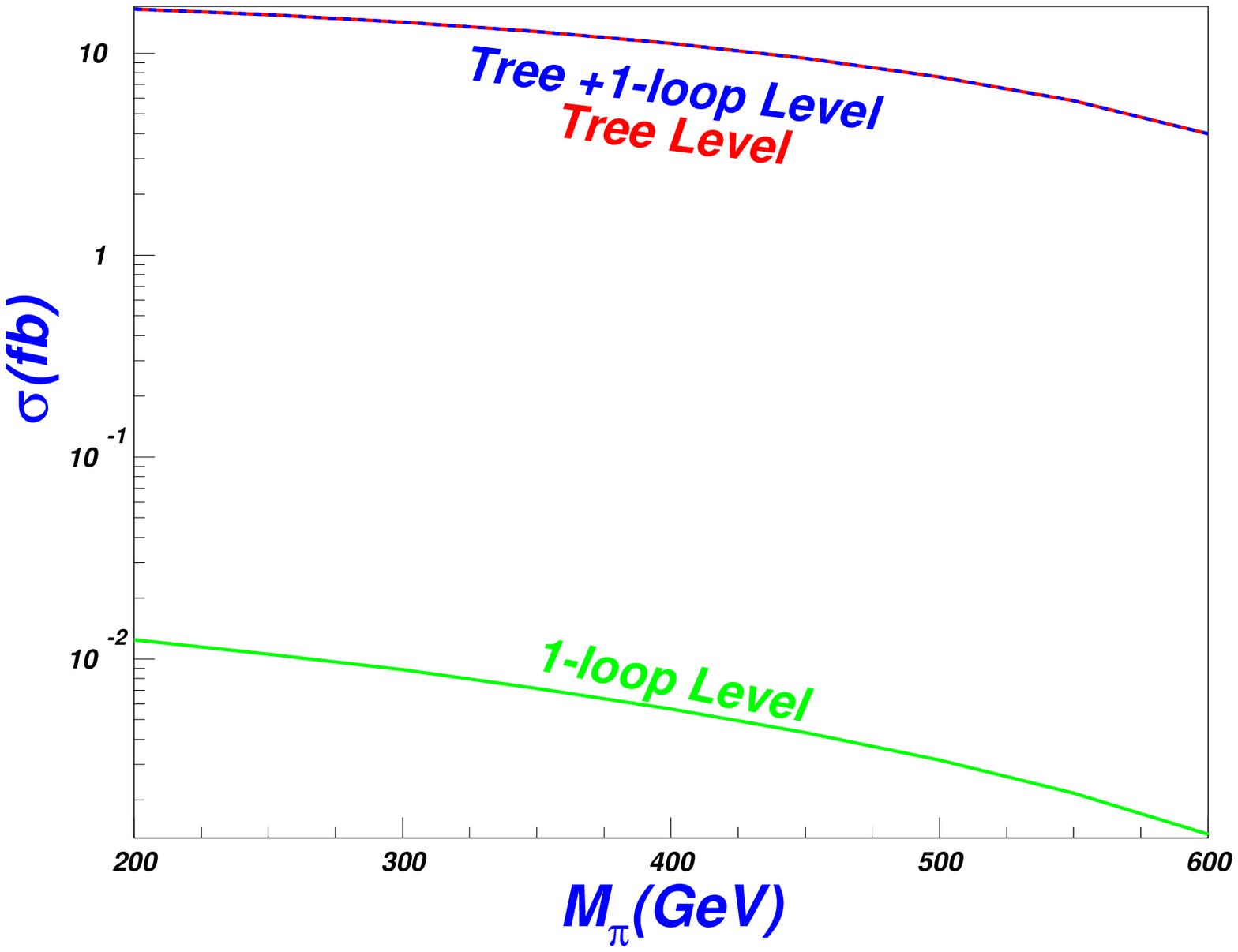,width=8.25cm} \figsubcap{a}}
 \parbox{8.05cm}{\epsfig{figure=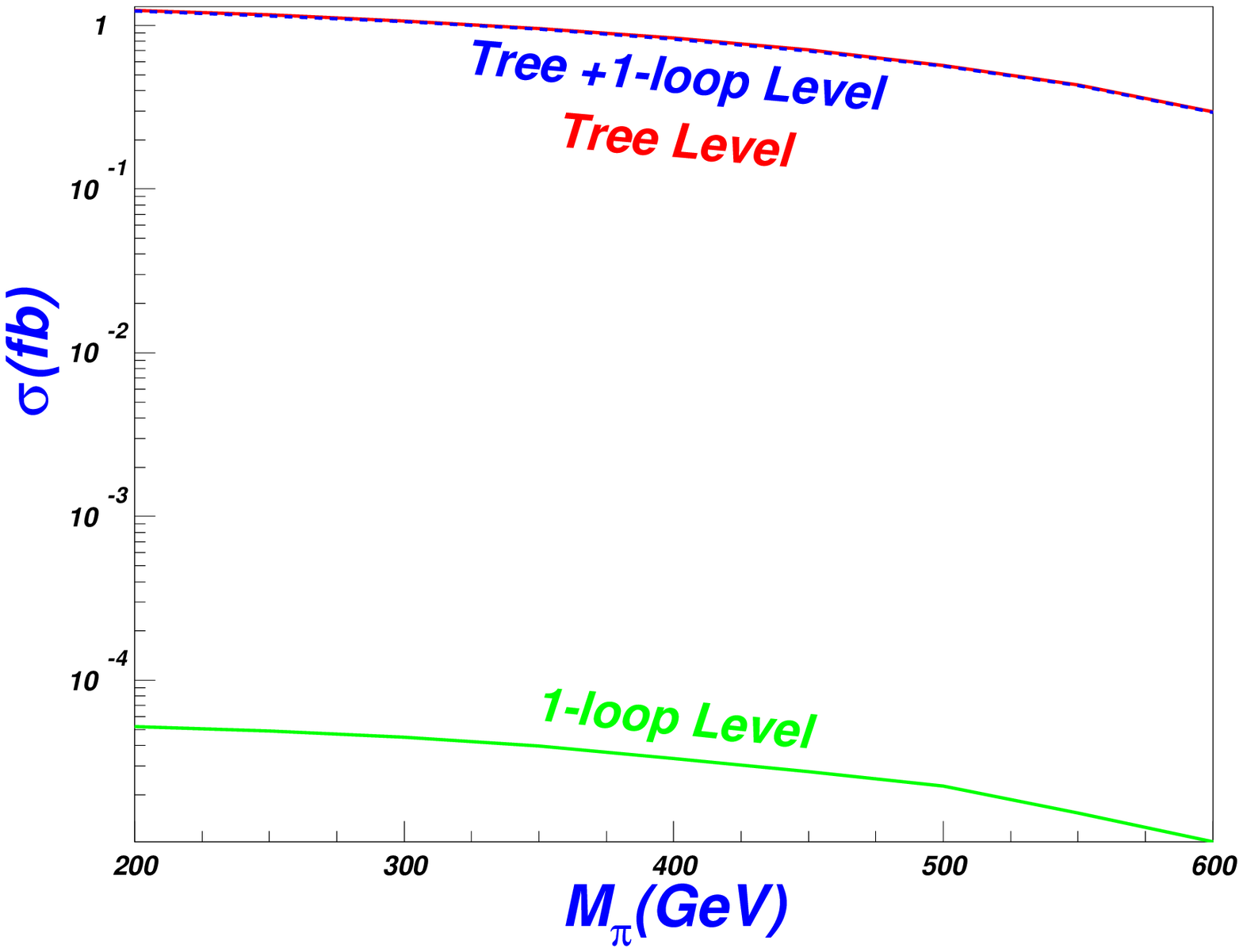,width=8.25cm} \figsubcap{b}}
     \caption{ Dependence of the cross section of  $ e^ + e^ - \to
\pi_t^+ \pi_t^- $ (a) and  $ e^ + e^ - \to \pi_t^0 h_t^0 $ (b) on
top-pion mass $m_\pi$ for $\sqrt{s} = 1500$ GeV. \label{eepp1} }
\end{center}
\end{figure}
The cross sections of the two productions $\pi_t^+ \pi_t^- $ and
$\pi_t^0 h_t^0 $ can be seen in Fig.\ref{eepp1}, from which, we can
see that the cross
 sections can reach $16.6$ fb and $1.24$ fb, for the charged and neutral
 production, respectively.  It agrees on our expectation that the neutral production
correction is smaller than that of the charged one. The reason is
twofold is that for the neutral scalar production $e^+e^- \to
\pi_t^0 h_t^0$, at the tree-level the photon does not contribute and
the $Z\pi_t^0 h_t^0$ couplings is almost the same as the $Z\pi_t^+
\pi_t^-$ coupling, which depresses the result. 

As for the one-loop level of the two processes, the contributions
are very small. The cross section of the charged pair production is
smaller than $0.012$ fb for $K_{UR}^{tc}=0.35$, even smaller of the
neutral production, less than $0.001$ fb.

  So we can estimate the contribution of the process
$e^+e^- \to \pi_t^0 h_t^0$ is smaller than that of the $e^+e^- \to
\pi_t^+ \pi_t^-$, which is verified in Fig.\ref{eepp1}(a)(b). The
interference between tree level and the one-loop level is also very
small, which can be seen that the cross section hardly change
whether we consider the one loop contribution or not.

Another thing is the parameter $K_{UR}^{tc}$ dependence. Since the
tree level contributions most and they are independent of the
$K_{UR}^{tc}$,  the cross section are almost the same with the
changing $K_{UR}^{tc}$.

Note that the identical productions $\pi_t^0\pi_t^0$ and
$h_t^0h_t^0$ are also considered and the rates are even small. Due
to the shortage of the tree level contribution, both processes are
proceeded at the one loop level. Moreover, the identical particles
in the final states add a $1/2$ factor, which even depress the cross
sections. That was verified by our calculation, the production rates
of the two identical processes, are less than $0.002$ fb in the
allowed parameter spaces. In view of the small contribution, we will
here not discuss them in detail.

\subsection{$\gamma\gamma \to
\pi_t^+\pi_t^-$  and $\gamma\gamma\to \pi_t^0 h_t^0$, $\pi_t^0
\pi_t^0$, $h_t^0 h_t^0$ at the PLC }

\def\figsubcap#1{\par\noindent\centering\footnotesize(#1)}
 \begin{figure}[bht]%
 \begin{center}
  \parbox{12.05cm}{\epsfig{figure=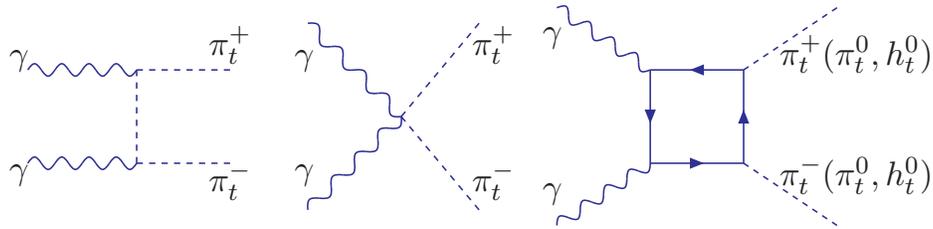,width=12.25cm}  }
 \caption{ The fenman diagrams of $\gamma\gamma \to \pi_t^+\pi_t^-$ and
 $\gamma\gamma \to \pi_t^0 h_t^0$ at the ILC.
\label{rrpp-fey} }
 \end{center}
 \end{figure}

\def\figsubcap#1{\par\noindent\centering\footnotesize(#1)}
\begin{figure}[bht]%
\begin{center}
\hspace{-0.8cm}
 \parbox{8.05cm}{\epsfig{figure=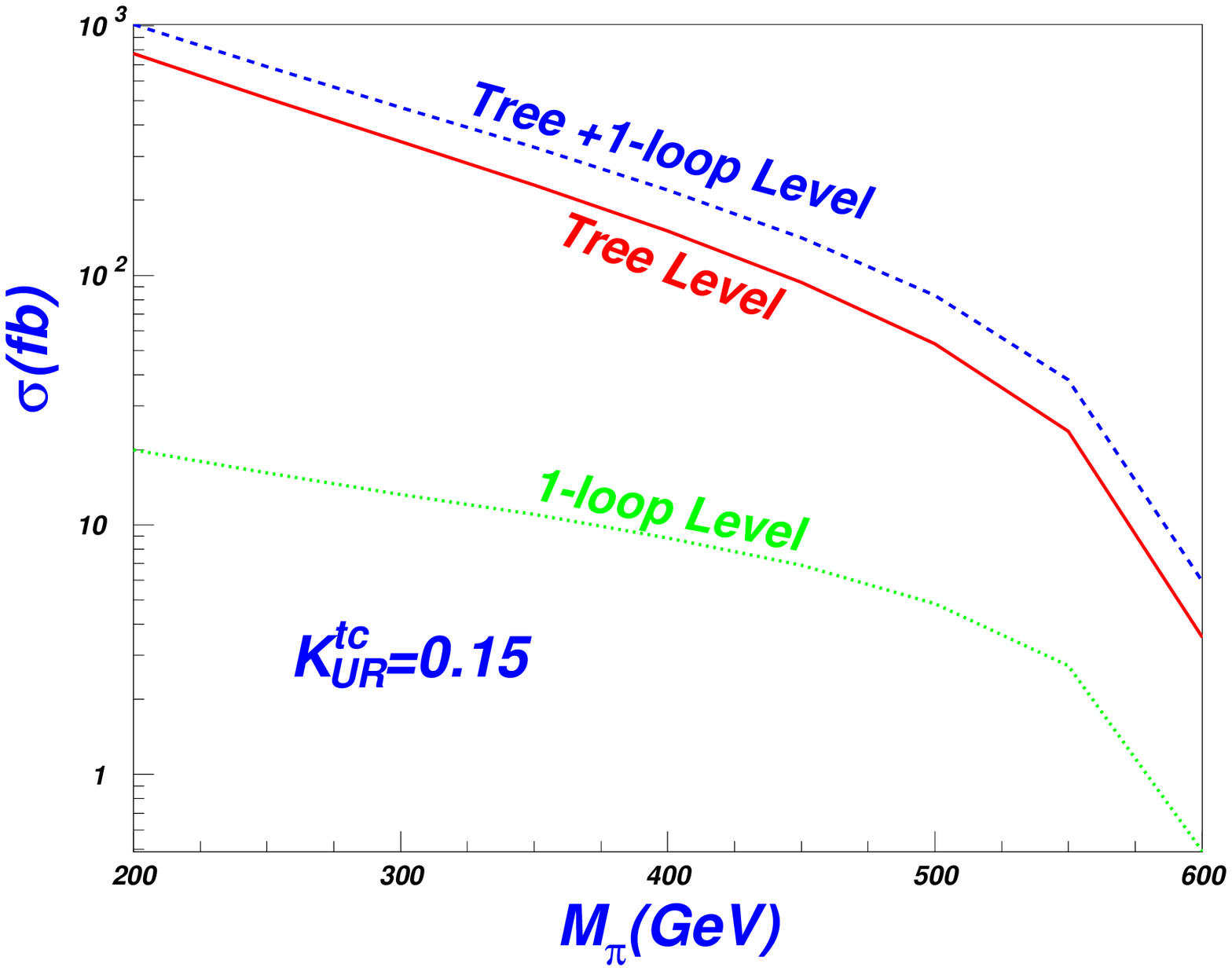,width=8.25cm} \figsubcap{a}}
 \parbox{8.05cm}{\epsfig{figure=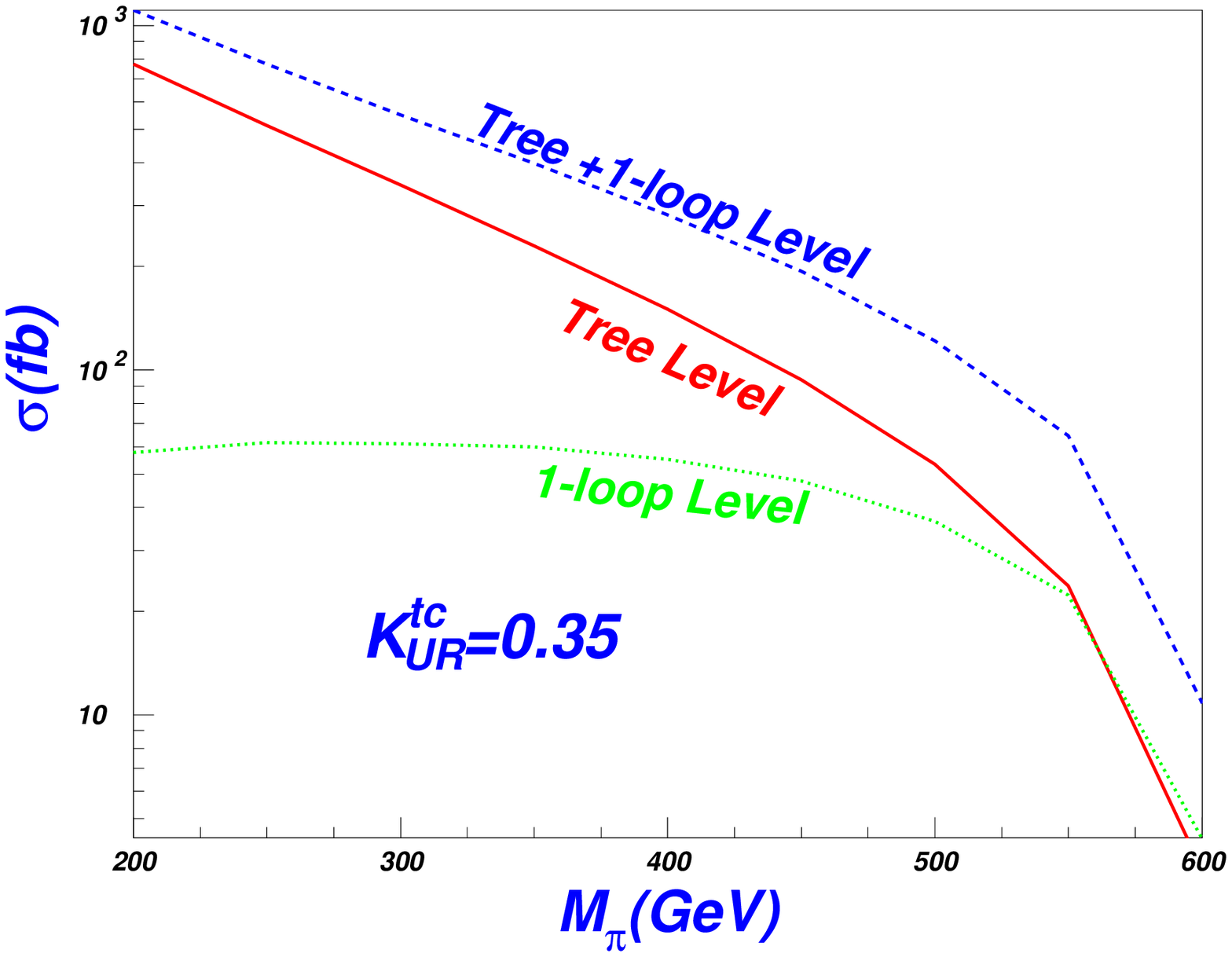,width=8.25cm} \figsubcap{b}}
     \caption{ Dependence of the cross section of  $ \gamma\gamma  - \to
\pi_t^+ \pi_t^- $  on top-pion mass $m_\pi$ with $\sqrt{s} = 1500$
GeV and for $K_{UR}^{tc}=0.15$ (a), ~$0.35$ (b). \label{rrpp1} }
\end{center}
\end{figure}

This processes carry out through by the $\gamma \pi_t^+ \pi_t^-$ and
$\gamma \gamma \pi_t^+ \pi_t^-$ couplings at the tree level,
$\pi_t^+ t \bar b$ coupling at the 1-loop level, just shown in
Fig.\ref{rrpp-fey}. Since there is no new effects restricted to the
TC2 model at the tree level, we also consider the 1-loop
corrections, which are consisted of the typical TC2 couplings and
one order smaller than the tree-level contribution, which can be
seen clearly in Fig.\ref{rrpp1}. Though the contributions may be
enhanced by different diagrams, the loop depression are
overwhelmingly larger so that the loop contribution is smaller that
of the tree level.

From Fig.\ref{rrpp1} we can also see the production rates can reach
one thousand fb and the cross sections decrease with the increasing
top-pion mass $m_\pi$ but larger than $1$ fb almost in all the
parameter space.

Compared the tree-level contribution of $e^+e^-\to \pi_t^+ \pi_t^-$
to that of  the $\gamma \gamma \to\pi_t^+ \pi_t^-$, we find that the
former is much smaller than the latter, the most important reason of
which is that, for the $e^+e^-\to \pi_t^+ \pi_t^-$, the contribution
is s-channel depression and the other process, i.e,  $\gamma \gamma
\to\pi_t^+ \pi_t^-$, is not infected with it.

\def\figsubcap#1{\par\noindent\centering\footnotesize(#1)}
\begin{figure}[bht]%
\begin{center}
\hspace{-0.8cm}
 \parbox{8.05cm}{\epsfig{figure=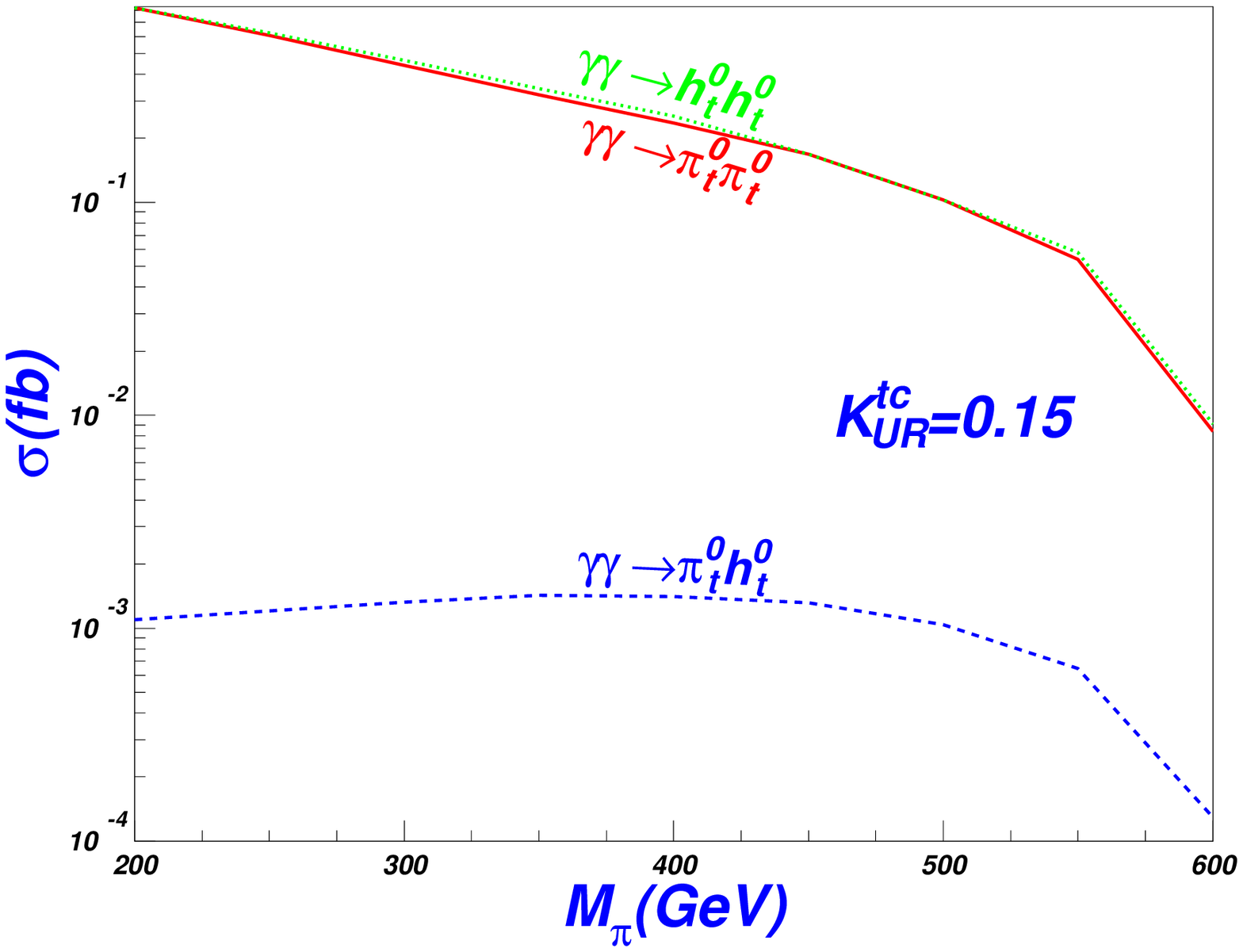,width=8.25cm} \figsubcap{a}}
 \parbox{8.05cm}{\epsfig{figure=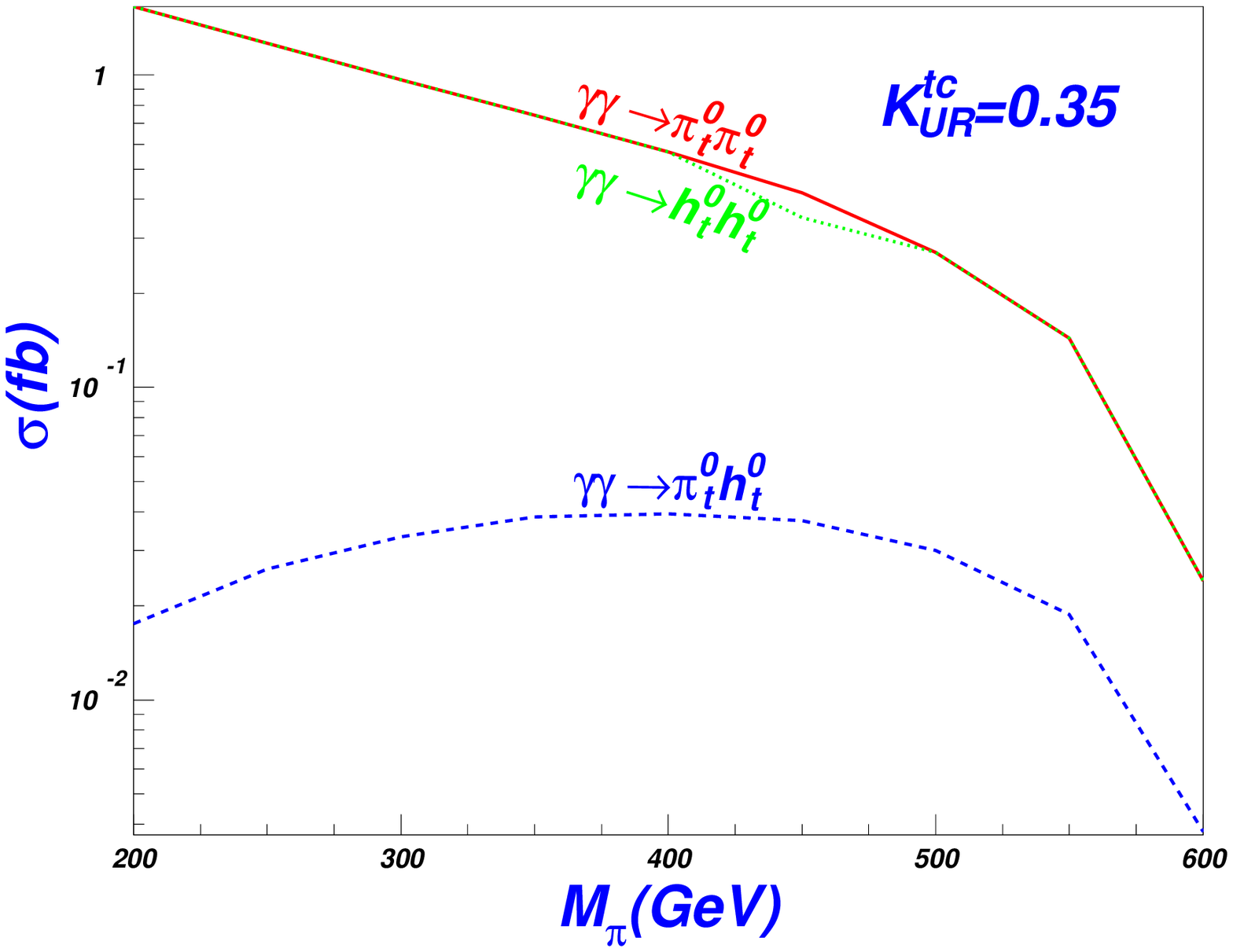,width=8.25cm} \figsubcap{b}}
     \caption{ Same as Fig.\ref{rrpp2}, but for $ \gamma\gamma \to
\pi_t^0 h_t^0 $, $ \gamma\gamma \to \pi_t^0 \pi_t^0 $ and $
\gamma\gamma \to h_t^0 h_t^0 $.  \label{rrpp2} }
\end{center}
\end{figure}

Since the photon can't couple to the neutral scalars directly, and
there is no tree level contribution, the neutral production process
carry through out in the one-loop, just as the last figure in
Fig.\ref{rrpp-fey}. Fig.\ref{rrpp2} shows the dependence of the
process $\gamma\gamma \to \pi_t^0h_t^0$, $\pi_t^0 \pi_t^0$, $h_t^0
h_t^0$ on the $m_\pi$ for $K^{tc}_{UR} =0.15$ and $0.35$,
respectively. We can see the cross section is smaller than $1$ fb in
quite a large parameter space and decrease with the increasing
$m_\pi$ from Fig.\ref{rrpp2}.

From Fig.\ref{rrpp1} and Fig.\ref{rrpp2}, we can also see that the
scalar pair productions don't vary too much as the $K^{tc}_{UR}$,
The reason is twofold. Firstly, At the tree level of charged
production, it is unconcerned about the parameter $K^{tc}_{UR}$.
Secondly, at the one-loop level of the charged and the scalar
productions, the $tttb$, $bbbt$ and $tttt$ contributes largely,
which are independent of $K^{tc}_{UR}$.

However, there is a little difference for the $\pi_t^oh_t^0$,  as
the $gg$ fusion processes.  Since the larger CP-even and the CP-odd
scalars cancel out,  the cross section may change largely with
varying  $K_{UR}^{tc}$.



\subsection{Simple Discussion of the ILC and the PLC Backgrounds Analysis }

Given the predictions listed in Fig.\ref{eepp1} and Fig.\ref{rrpp1},
we now discuss their observability at the ILC. Firstly, for the
 final state $\pi_t^+ \pi_t^-$, the
charged top-pions $\pi_t^+$ still decays to $t\bar b$ with the
branching ratio of $70\%$ and the signal is the same as that at the
LHC, i.e. the $2b+2j+2l+\E_slash$ signal. At the same time, for the
$\pi_t^0 h_t^0$ final states, the neutral scalars decay to $t\bar
c$, and the signal is also $2b+2j+2l+\E_slash$, the same as the
charged production.

So for the two final states, the mainly SM ILC backgrounds will be
$e^+e^- \to WWhh$, $WWZZ$, $WWZjj$, $t\bar tW$, $t\bar tjj$, the
cross sections of the processes $WWhh$, $WWZZ$, $WWZjj$, $t\bar tW$,
however, are below the order of $1$ fb. The cross section of the
last one, i,e, the $t\bar tjj$ production in the $e^+e^-$ collision,
are quite large, about 10fb.

 Similarly, for the $\gamma\gamma$ collisions of the same final states $\pi_t^+
\pi_t^-$ and  $\pi_t^0 h_t^0$, the main SM backgrounds is
$\gamma\gamma \to WWhh$, $WWZZ$, $WWZjj$ , $t\bar tW$, $t\bar tjj$,
in which the cross section of the last one can arrive at $30$ fb.

The backgrounds are in the same level of the signal, so we have to consider the depression.
If the transverse momenta cuts, e.g. $p_t^j>20$ GeV
and the b-tagging skills,  with $60\%$ b-tagging efficiency and $1\%$ mistagging, are employed, the
backgrounds may be depressed very largely. Moreover, if the top-pion masses are reconstructed, the
signal should be chosen out more clearly.


So we here assume that the pair PGB production at the ILC and the
$\gamma\gamma$ collisions with a cross section larger than $10$ fb
may be observable at $95\%$ C.L. for an integrated luminosity of
$100$ fb$^{-1}$. Compared with the predictions in Fig.\ref{eepp1}
and Fig.\ref{rrpp1}, one sees that TC2 model can enhance the
production $e^+e^- \to \pi_t^+\pi_t^-$ and $\gamma\gamma \to
\pi_t^+\pi_t^-$ and may be observable at the ILC in a large part of
the parameter space.

\section{summary and conclusion  }

We considered the PGB pair productions in the TC2 model, proceeding
through $gg\to \pi_t^+\pi_t^-$, $gg\to \pi_t^0 h_t^0$, $\pi_t^0
\pi_t^0$, $h_t^0h_t^0$, $q \bar{q} \to \pi_t^+\pi_t^-$, $q \bar{q}
\to \pi_t^0 h_t^0$, $\pi_t^0 \pi_t^0$, $h_t^0h_t^0$, $e^+e^-\to
\pi_t^+\pi_t^-$, $e^+e^-\to \pi_t^0 h_t^0$,  $\gamma\gamma\to
\pi_t^+\pi_t^-$, and $\gamma\gamma\to \pi_t^+\pi_t^-$, $\pi_t^0
\pi_t^0$, $h_t^0h_t^0$ as a probe of the TC2 model. Since the
backgrounds can be effectively suppressed by the scalar mass
reconstruction, these processes can be used to probe the TC2 model.
We found that these PGB pair productions in different collisions can
play complementary roles in probing the TC2 model:

At the LHC, the cross section is large, and firstly we discussion
the rates at the parton level, one by one,  to compare their
relative contribution.
\begin{itemize}
\item[{\rm (1)}] For the $\pi_t^+\pi_t^-$ production at the LHC, the
processes  may be detectable when the cross sections reach $1000$
fb, as discussed in the above section. For $gg\to \pi_t^+\pi_t^-$,
the cross section can reach $1000$ fb in most of the parameter
spaces, which contributes large for this charged production.
 For $b\bar b\to \pi_t^+\pi_t^-$ and $c\bar c\to
\pi_t^+\pi_t^-$, the cross sections can arrive at $1000$ fb in most
of the parameter spaces, which are also large at the LHC. For $u\bar
u\to \pi_t^+\pi_t^-$, the cross section decreases with increasing
$m_\pi$ rapidly, though it is about $200$ fb when $m_\pi=200$ GeV.
When $m_\pi=400$ GeV, the cross section is only $13$ fb. As the
cross sections for other processes, $d\bar d\to \pi_t^+\pi_t^-$ and
$s\bar s\to \pi_t^+\pi_t^-$ are even smaller. So the processes
$u\bar u,~d\bar d,~s\bar s  \to \pi_t^+\pi_t^- $ can't contribute
much at the LHC in most of the parameter space.

\item[{\rm (2)}]The process $pp\to \pi_t^0 \pi_t^0$ and $pp\to h_t^0
h_t^0$ can arrive $2000$ fb for $\sqrt{s}=14$ TeV, but with
increasing $m_\pi$ and decreasing $\sqrt{s}$, the production rate
decreases rapidly. For example, when $\sqrt{s}=7$ TeV and
$m_\pi=600$ GeV, the productions are only about $0.5$ fb.

\item[{\rm (3)}]The process $gg\to \pi_t^0 h_t^0$ is closely connected to the
parameter $K_{UR}^{tc}$. The cross section, however, is not too
large.  The cross section With $K_{UR}^{tc}=0.35$ and $\sqrt{s}=14$
TeV, for example, is smaller than $430$ fb. The $q\bar q \to \pi_t^0
h_t^0$ is much smaller.

\item[{\rm (4)}] For the process $u \bar d \to \pi_t^+ \pi_t^0$, the
cross section is larger than $100$ fb in most of the parameter space, while
$c \bar s \to \pi_t^+ \pi_t^0$,
smaller than $100$ fb in quite a large space, so the parton production of the
 $u \bar d \to \pi_t^+ \pi_t^0$ contributes
most in the $pp \to \pi_t^+ \pi_t^0$ process.
\end{itemize}

\def\figsubcap#1{\par\noindent\centering\footnotesize(#1)}
\begin{figure}[bht]%
\begin{center}
\hspace{-0.8cm}
 \parbox{8.05cm}{\epsfig{figure=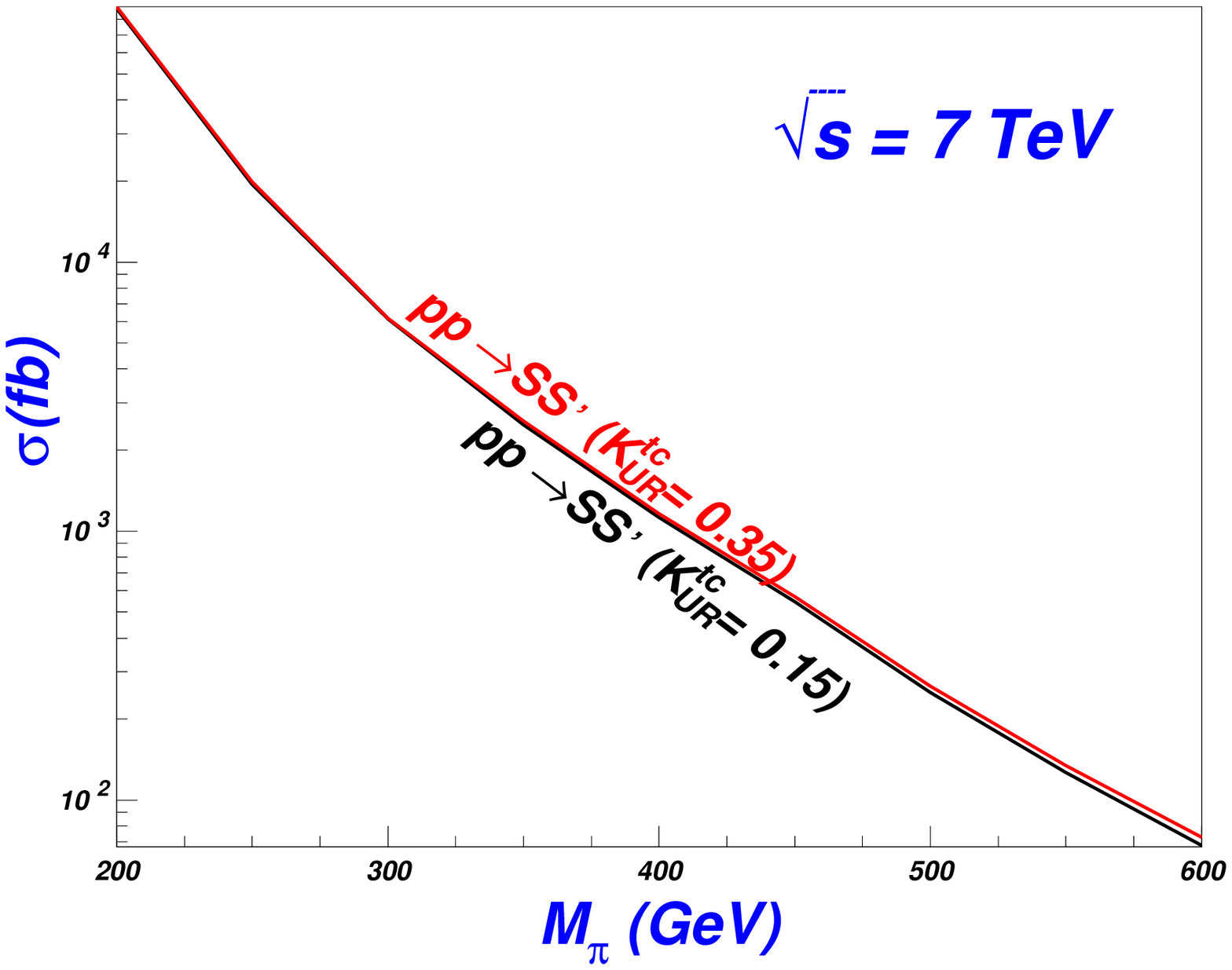,width=8.25cm}
 \figsubcap{a}}
 \hspace*{0.2cm}
 \parbox{8.05cm}{\epsfig{figure=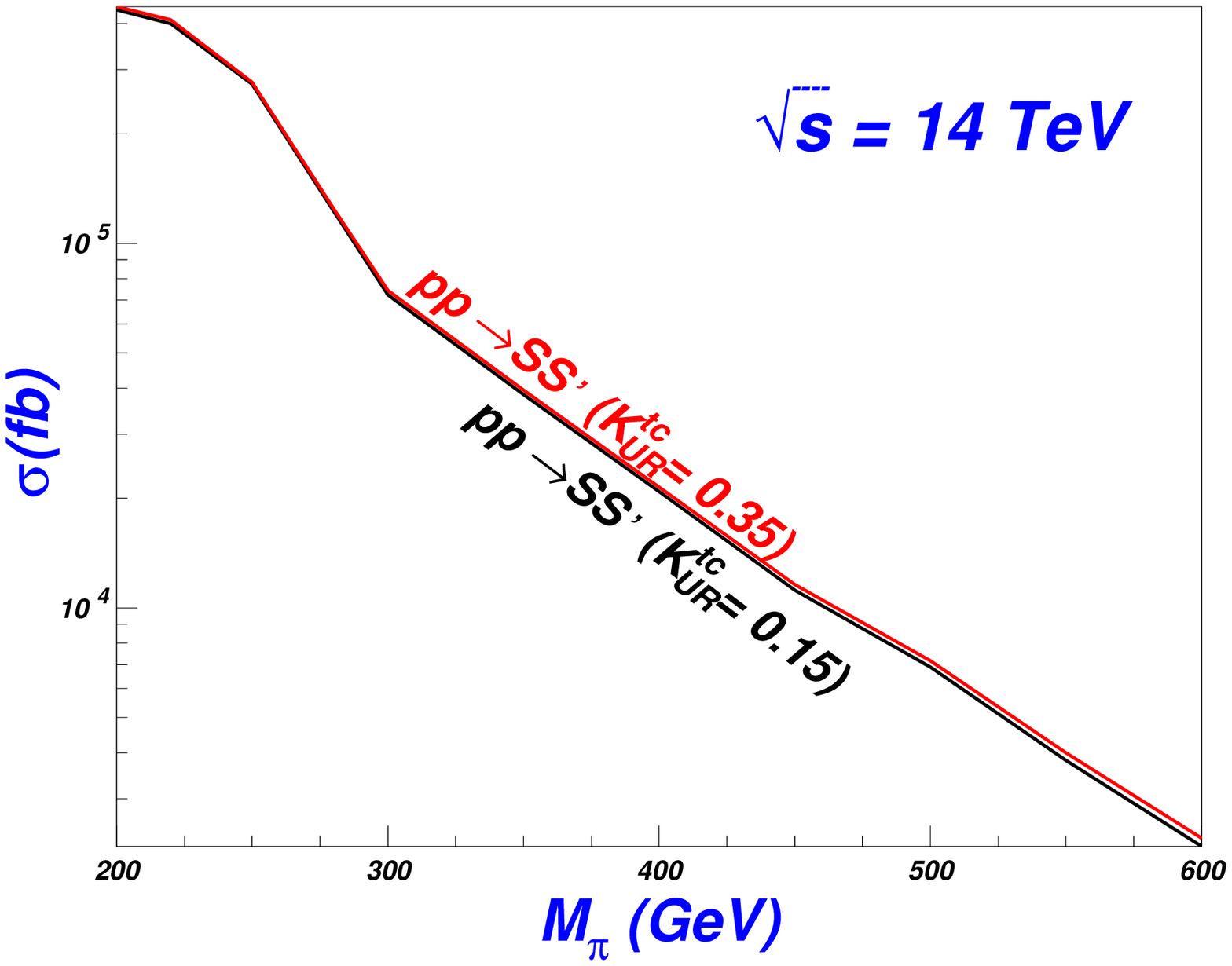,width=8.25cm}
 \figsubcap{b}}
 \caption{ The total cross section $\sigma$ of the processes
 $p\bar p\to SS' $ as a function of the top-pion mass $m_{\pi_t}$ with
 $\sqrt{s}=7$ TeV and $\sqrt{s}=14$ TeV and for
$K_{UR}^{tc}=0.15, ~ 0.35$. \label{total} }
\end{center}
\end{figure}


The total cross sections $\sigma$ (with all the initial states
contribution summed and the final states summed)
  of the processes
 $p\bar p\to SS' $ as a function of the top-pion mass $m_{\pi_t}$ with
 $\sqrt{s}=7$ TeV and $\sqrt{s}=14$ TeV and for
$K_{UR}^{tc}=0.15, ~ 0.35$ are given in Fig.\ref{total}. We can see from Fig.\ref{total} that the
total cross sections decrease
 with the increasing top pion mass, and vary very slightly with the parameter $K_{UR}^{tc}$,
 since the charged production is dominant, which is free from the $K_{UR}^{tc}$. We
 can also see that the cross sections are quite large, about $1000$ fb for $\sqrt{s} = 7$ TeV in a good case, much
 larger than $1000$ fb for $\sqrt{s} = 14$ TeV  in a large part.

 According to the background analysis above, the total production $pp\to SS'$ may be detected
at the LHC with quite a large possibility.

For the $\pi_t^+\pi_t^-$ production at the ILC, the processes may be
detectable when the cross sections reach $10$ fb. For $e^+e^-\to
\pi_t^+\pi_t^-$ and $\gamma\gamma\to \pi_t^+\pi_t^-$, the cross
section can reach $20$ fb in most of the parameter spaces, which is
possible to be detected at the ILC. Similarly, the processes
$e^+e^-\to \pi_t^0 h_t^0$ and $\gamma\gamma\to \pi_t^0 h_t^0$, below
$10$ fb in most of the parameter space, difficult to observe at the
ILC if the signal are singled out. According to discussion above,
however, the charged and the neutral productions have the same
collider signature, so the total cross section at the ILC and the
PLC may be both observable.

As a conclusion, as long as the top-pions are not too heavy, e.g.,
below $500$ GeV, the productions might be detectable at the LHC, the
ILC and the PLC. In general, the charged pion pair productions have
larger possibility to be detectde since their couplings to $t \bar
b$ are not suppressed by $K_{UR}^{tc}$. while for the neutral pion
productions, the cross sections are closely connected to the
parameter $K_{UR}^{tc}$.
We see from the figures listed above that in a large part of the
parameter space the cross sections of the scalar pair productions
can reach the possible detectable level ($1000$  fb for the LHC and
$10$ fb for the ILC). Therefore, the pair productions of PGBs may
serve as a good probe of the TC2 model.

\section*{Acknowledgments}

We would like to thank J. J. Cao, Q. Yan, J. M. Yang, and S. Yang for
helpful discussions. This work was supported by the National Natural Science
Foundation of China under the Grants No.11105125.


\end{document}